\documentclass[aps,nofootinbib,prd,superscriptaddress,twocolumn,10pt]{revtex4}

\usepackage{color,amsmath,amssymb,graphicx,latexsym,txfonts}
\usepackage{threeparttable,multirow,subfigure,booktabs,lineno,tabularx,ulem}

\newcommand{\nn}{{\nonumber}}

\newcommand{\beq}{\begin{equation}}
\newcommand{\eeq}{\end{equation}}
\newcommand{\bea}{\begin{eqnarray}}
\newcommand{\eea}{\end{eqnarray}}

\newcommand{\gsim}{\lower.7ex\hbox{$\;\stackrel{\textstyle>}{\sim}\;$}}
\newcommand{\lsim}{\lower.7ex\hbox{$\;\stackrel{\textstyle<}{\sim}\;$}}

\newcommand{\be}{\begin{equation}}
\newcommand{\ee}{\end{equation}}
\newcommand{\ba}{\begin{eqnarray}}
\newcommand{\ea}{\end{eqnarray}}

\begin{document}

\title{Earth shielding and daily modulation from electrophilic boosted dark particles}
\author{Yifan Chen\footnote{yifan.chen@nbi.ku.dk}}
\affiliation{Niels Bohr International Academy, Niels Bohr Institute, Blegdamsvej 17, 2100 Copenhagen, Denmark}
\affiliation{CAS Key Laboratory of Theoretical Physics, Institute of Theoretical Physics, Chinese Academy of Sciences, Beijing 100190, China}

\author{Bartosz Fornal\footnote{bfornal@barry.edu}}
\affiliation{Department of Chemistry and Physics, Barry University, Miami Shores, Florida 33161, USA}

\author{Pearl Sandick\footnote{sandick@physics.utah.edu}}
\affiliation{Department of Physics and Astronomy,
University of Utah, Salt Lake City, Utah 84112, USA}

\author{Jing Shu\footnote{jshu@pku.edu.cn}}
\affiliation{CAS Key Laboratory of Theoretical Physics, Institute of Theoretical Physics, Chinese Academy of Sciences, Beijing 100190, China}
\affiliation{School of Physical Sciences, University of Chinese Academy of Sciences, Beijing 100049, China}
\affiliation{School of Fundamental Physics and Mathematical Sciences, Hangzhou Institute for Advanced Study, University of Chinese Academy of Sciences, Hangzhou 310024, China}
\affiliation{International Center for Theoretical Physics Asia-Pacific, Beijing/Hangzhou, China}

\author{Xiao Xue\footnote{xiao.xue@desy.de}}
\affiliation{II. Institute of Theoretical Physics, Universit\"{a}t  Hamburg, 22761, Hamburg, Germany}
\affiliation{Deutsches Elektronen-Synchrotron DESY, Notkestr. 85, 22607, Hamburg, Germany}

\author{Yue Zhao\footnote{zhaoyue@physics.utah.edu}}
\affiliation{Department of Physics and Astronomy,
University of Utah, Salt Lake City, Utah 84112, USA}

\author{Junchao Zong\footnote{jczong@smail.nju.edu.cn}}
\affiliation{Department of Physics, Nanjing University, Nanjing 210093, China}
\affiliation{CAS Key Laboratory of Theoretical Physics, Institute of Theoretical Physics, Chinese Academy of Sciences, Beijing 100190, China}

\begin{abstract}
 \vspace{1mm}
Boosted dark particles of astrophysical origin  can lead to nonstandard nuclear or electron recoil signals in direct detection experiments. 
We conduct an investigation of the daily
 modulation feature of a potential future  signal of this type. In particular, we perform   simulations of the dark particle  interactions with  electrons in atoms building up the Earth on its path to the detector and provide in-depth predictions for the expected daily  changes in the signal for various direct detection experiments, including XENONnT, PandaX, and LUX-ZEPLIN.
\end{abstract}
\maketitle

\section{Introduction}

Dark matter (DM) is certainly one of the greatest outstanding puzzles in modern particle physics. An enormous scientific effort has been undertaken, both on the theoretical and experimental sides, to shed more light on its nature, with  great progress achieved in probing the parameter space of various particle physics models. Direct detection experiments offer a particularly promising way to search for DM, since in many models the DM particle is expected to undergo measurable recoils of nuclei and/or electrons in the detector. 

A signal of this type has been hinted by the XENON1T experiment, which observed an excess of low-energy electron recoil events  \cite{XENON:2020rca}. One of the beyond Standard Model (SM) interpretations  involved a boosted dark matter particle scattering on electrons \cite{Fornal:2020npv,Kannike:2020agf}. 
Although this  effect was recently ruled out by the  XENONnT experiment \cite{XENONCollaboration:2022kmb}, and was most likely the result of beta decays of tritium, it still remains interesting to explore the possibility of detecting such signals in the future in XENONnT itself and other experiments.

To keep our analysis general, we will not require for the boosted particle to be the DM, thus we will refer to it simply as a boosted dark particle (BDP).
If the BDP  $\chi$ is much heavier than the electron, the observed electron
energy deposition signal implies $\chi$ velocities of $\mathcal{O}(10^{-1})\,c$. 
Such fast-moving BDPs cannot come from the Milky Way halo and, instead, must be of  astrophysical origin, produced, e.g., via semiannihilation $\bar\chi +\bar\chi \to {\chi}+ X$ (where $X$ is a SM particle or a new particle  eventually decaying to SM particles)~\cite{DEramo:2010keq}, or via annihilation of a heavier dark sector particle, $\psi$, $\psi+\bar\psi\to \chi+\bar\chi$~\cite{Boehm:2003ha}. 
Either the Galactic Center (GC) or the Sun can be the dominant source of the BDP flux.
In \cite{Berger:2014sqa}  specific particle physics models for $\chi$ and $\psi$ are discussed, and $\psi$ is shown to satisfy the requirements  for a DM candidate, including an annihilation cross section leading to the observed DM relic density.


Searches for such BDPs have been proposed for
 large volume neutrino experiments, e.g., Super-Kamiokande \cite{Huang:2013xfa,Agashe:2014yua,Berger:2014sqa,Kachulis:2017nci}, ProtoDUNE \cite{Chatterjee:2018mej,Kim:2018veo}, IceCube \cite{Bhattacharya:2016tma,Kopp:2015bfa}, and DUNE \cite{Berger:2019ttc,Arguelles:2019xgp,Abi:2020evt,Kim:2019had} (see also  \cite{Kamada:2017gfc,Kamada:2018hte,McKeen:2018pbb,Kamada:2019wjo} for related work). Here we focus on electrophilic BDPs, and our results are applicable to direct detection experiments like  XENONnT, PandaX \cite{PandaX:2014mem} and LUX-ZEPLIN \cite{LZ:2022ufs}.
For electrophilic BDPs, when the BDP-electron scattering cross section is sizable, the electron ionization signal in direct detection experiments is expected to exhibit daily modulation due to the Earth shielding effect \cite{Fornal:2020npv}. This can be used to distinguish the BDP signal from various backgrounds.  The information on the phase of the modulation can reveal the direction of the BDP flux, which would be of high importance for experimental analyses. 

In this paper we 
 extend the analysis of the daily modulation of the BDP signal of astrophysical origin, and  explicitly account for the distribution of various elements inside the Earth, calculating their contributions to the BDP-electron ionization cross section. Our results apply to any direct detection experiment measuring electron recoil energies. 
The software used in this research is publicly available. The ionization form factor is calculated using {\tt AtomIonCalc}\footnote{https://github.com/XueXiao-Physics/AtomIonCalc}, which is refined from the software {\tt DarkARC} \cite{Catena:2019gfa,DarkARC}. The software {\tt realEarthScatterDM}\footnote{https://github.com/XueXiao-Physics/realEarthScatterDM} is used to simulate the BDP propagation inside the Earth, and was independently developed for this research.

\section{Boosted Dark Particle Model and Sources}
The particle physics model we consider here  is a simple extension of the SM including four new particles: the DM $\psi$, the BDP $\chi$, and the dark mediators $V$ and $Z'$.  The dark mediator $Z'$ enables the annihilation $\psi+\bar\psi\to \chi+\bar\chi$, the cross section for which is given by \cite{Berger:2014sqa}
\bea
&&\hspace{-6mm}\langle\sigma_{\chi\bar\chi} v\rangle = \frac{{g'}_{\!\psi}^{2} {g'}_{\!\chi}^2}{12 \pi\,m_\psi} \frac{\sqrt{m_\psi^2 - m_\chi^2}}{(4m_\psi^2-m_{Z'}^2)^2 + \Gamma^2_{Z'}m_{Z'}^2}\nonumber\\ &\times&\left[3m_\chi^2 + v^2 \, \frac{24m_\psi^4 m_\chi^2+ m_\psi^2 m_{Z'}^2(m_{Z'}^2 - 6 m_\chi^2)-m_\chi^2m_{Z'}^4}{m_{Z'}^4}\right], \ \ \ \ 
\eea
where $g'_\psi$ and $g'_\chi$ are the couplings of $\psi$ and $\chi$ to $Z'$, respectively, and $\Gamma_{Z'}$  is the width of the $Z'$. 
There is a wide range of parameter values for which this yields the correct DM relic density, i.e., when 
${\langle\sigma_{\chi\bar\chi} v\rangle} \approx 3\times 10^{-26} \ {\rm cm^3/s}$. However, there can also exist other DM annihilation channels involving SM particles, in which case ${\langle\sigma_{\chi\bar\chi} v\rangle}$ can be much smaller.
Some representative benchmark points are provided in \cite{Berger:2014sqa}.

The
BDP interacts with electrons through the dark mediator $V$, as described by the
Lagrangian terms 
\bea\label{111}
\mathcal{L} \supset g_\chi V_\mu\bar\chi\gamma^\mu\chi + g_e V_\mu\bar{e}\gamma^\mu e \ .
\eea
If the mediator mass $m_V$ is much larger than the $\sim {\rm keV}$ momentum transfer, in the parameter space of interest the cross section for BDPs scattering on free electrons  simplifies to
\bea
\overline{\sigma_e} = \frac{g_\chi^2g_e^2 m_e^2}{\pi\, {m_V^4}} \ .
\eea
The benchmark points we consider here correspond to 
 BDP-electron scattering cross sections of $\overline{\sigma_e} = 10^{-28} \ {\rm cm}^2$, $10^{-31} \ {\rm cm}^2$, and  $10^{-33} \ {\rm cm}^2$.  In those scenarios, there is a wide range of values for the parameters $g_\chi$, $g_e$ and $m_V$  consistent with existing experimental bounds (see \cite{Fornal:2020npv} for details). 

On the other hand, an electrophilic BDP can couple to a proton at the loop level through a mixing induced by charged leptons \cite{Kopp:2009et}. However, we will not consider the resulting BDP-nucleus scattering channel for two reasons. First, the ionization cross section for the BDP, which is similar to an elastic scattering process on an electron at rest, typically dominates over the scattering cross section on a nucleus for $m_\chi$ below $100$ MeV.  If we take oxygen as a benchmark, which dominates  the inside of the Earth, the ratio is approximately $28$ for $m_\chi = 100$ MeV. Furthermore, the coupling between a leptophilic BDP and a nucleon can be more suppressed if the massive mediator between the BDP and the lepton is a scalar or an axial vector. As discussed in \cite{Kopp:2009et}, while maintaining the same parametric dependence in the BDP-electron scattering cross section, the BDP-proton interaction is only introduced at the 2-loop level for the scalar mediator, and it is absent up to 2-loops for the axial-vector mediator. 
Since including the BDP-nucleus scattering would not qualitatively change our results, 
we will neglect it.

The two main candidates for sources of a BDP flux are DM annihilation in the Galactic Center (GC) or halo, and DM capture with subsequent annihilation inside the Sun. 
Regarding the first possibility, the expected full-sky BDP flux from the GC can be estimated as 
 \cite{Agashe:2014yua} 
\begin{eqnarray}\label{two-comp}
\Phi^{\rm BDP}_{\rm GC} &\approx& \frac{1.6\times 10^{-2}}{{\rm cm}^{2}\,{\rm s}}\,\bigg(\frac{\langle\sigma_{\chi\bar\chi} v\rangle}{5\times 10^{-26}\,{\rm cm}^3{\rm s}^{-1}}\bigg) \,\bigg(\frac{100 \, {\rm MeV}}{m_{{\chi}}}\bigg)^2 \ , \ \ \ \ 
\end{eqnarray}
where  $\langle\sigma_{\chi \bar\chi} v\rangle$ denotes the  thermally averaged annihilation cross section for $\psi + \bar\psi \to \chi + \bar\chi$. 
For example, assuming 
thermally produced DM with mass $m_\chi = 100 \,\rm MeV$, the expected BDP flux is  $\Phi^{\rm BDP}_{\rm GC} \lesssim 1.6\times 10^{-2} \, {\rm cm}^{-2}{\rm s}^{-1}$,  saturated when $\psi + \bar\psi \to \chi + \bar\chi$ is the only annihilation channel,  otherwise smaller.
If the DM is produced nonthermally, then the annihilation cross 
 section can be larger. 

As for the second possibility, if the DM particles scatter off nuclei, they can be captured by the Sun and accumulate in its core. 
As discussed in \cite{Berger:2014sqa}, the Sun reaches capture-annihilation equilibrium for typical values of the  DM scattering and annihilation cross sections, and the BDP flux becomes fully determined by the DM capture rate, taking the form  \cite{Agashe:2014yua}\footnote{In Eq.\,(\ref{PhiBDPsun}) we ignored the effects of the BDPs scattering off electrons inside the Sun. However, given that the typical electron energies  in the Sun are
 $\sim$ keV, their 
interaction with BDPs are not expected to attenuate the flux by much. For a more detailed discussion of this effect, see \cite{An:2017ojc}.}
\begin{eqnarray}\label{PhiBDPsun}
\Phi^{\rm BDP}_{\rm Sun} &\approx& \frac{7.2\times 10^{-2}}{{\rm cm}^{2} \,\rm s}\,\bigg(\frac{\sigma_{\rm nucl}}{10^{-42} \, {\rm cm}^2}\bigg)\,\bigg(\frac{100 \, {\rm MeV}}{m_{\chi}}\bigg)^2\ ,
\end{eqnarray}
where $\sigma_{\rm nucl} \sim 10^{-42} \,{\rm cm}^2$ is on the order of the upper limit for the DM-nucleon scattering cross section set by spin-dependent DM direct detection experiments \cite{XENON:2019rxp}, and it was assumed that, at leading order, the scattering cross section is velocity-independent\footnote{The DM capture rate can be enhanced by a factor of up to $\sim25$ if the
leading order cross section has a $v^2$ dependence.}.

\section{Model of the Earth}\label{sec_MoE}

\begin{figure}[t!]
  	\centering
  	\includegraphics[width = 1 \columnwidth]{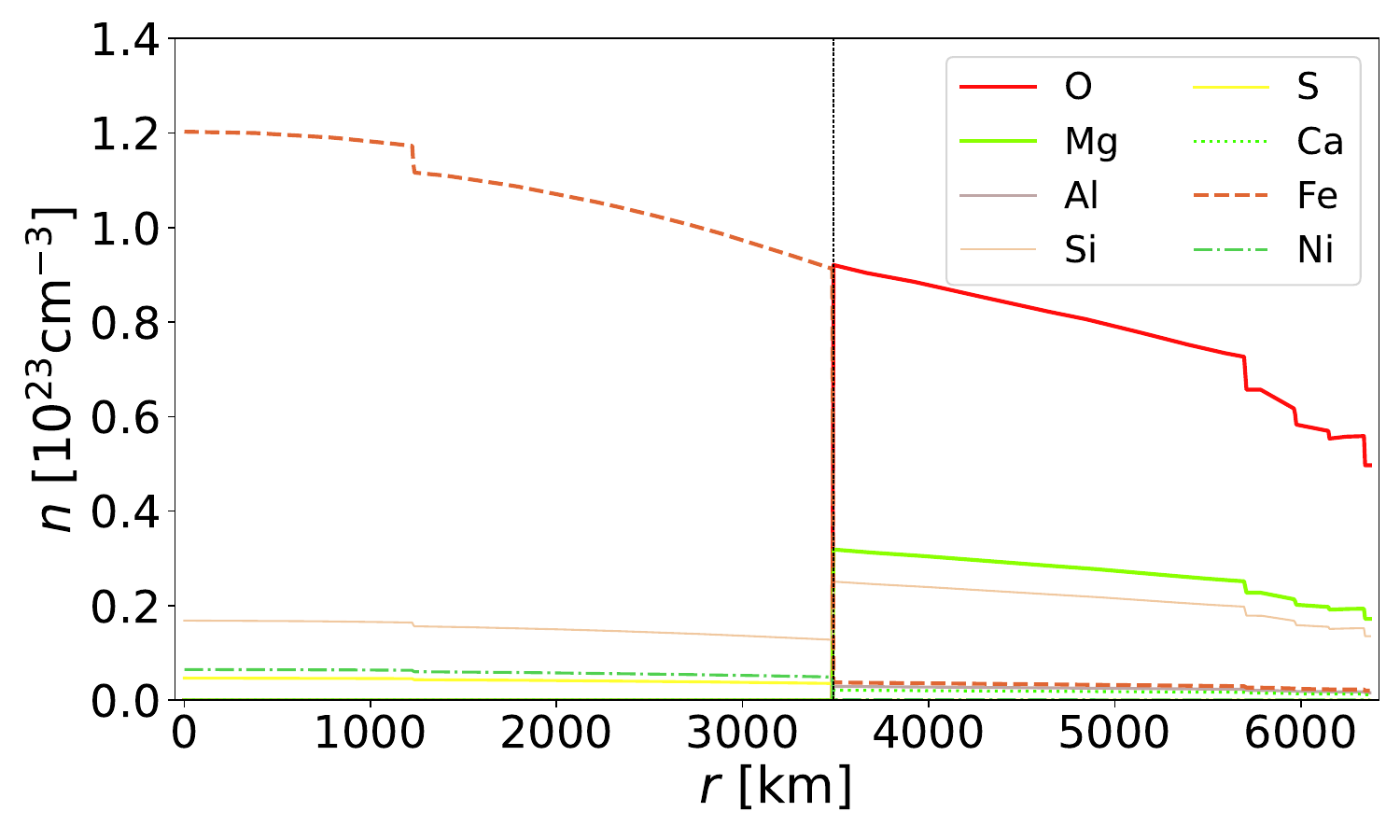}
  	\caption{
  		{Number density of various atoms in the Earth's core and mantle. The mantle-core border is indicated by a vertical black dotted line. The density profile was taken from \cite{EarthDensityProfile}, while the Earth's composition was adopted from  \cite{EarthComposition}.}}
  	\label{nCM}
\end{figure}

The Earth consists essentially  of two parts: the core and the mantle. 
The eight most abundant atomic elements in the core and mantle \cite{EarthComposition,Kavanagh:2016pyr}  are shown in Table \ref{EarthModel}. The remaining elements contribute a mass fraction below $1\%$.  Due to the lack of precise information regarding the density of each element in terms of the distance  from the Earth's center, we assume that a given element's mass fraction is constant in the core and mantle, and we take the value in each region to be the average value in Table \ref{EarthModel}. 
The total density profile as a function of radius is taken from \cite{EarthDensityProfile} and is shown in Fig.\,\ref{nCM}. The Earth is assumed to be isotropic, despite the complexity of its composition.

\begin{table}[t!]
\begin{center}
\begin{tabularx}{\columnwidth}{XccXX}
\hline \hline Element & $Z$ & \(m_{A}\,[\mathrm{GeV}]\) & Core & Mantle \\
\hline Oxygen, O & 8 & 14.9 & 0.0 & 0.440 \\
Magnesium, Mg & 12 & 22.3 & 0.0 & 0.228 \\
Aluminium, Al & 13 & 25.1 & 0.0 & 0.0235 \\
Silicon, Si & 14 & 26.1 & 0.06 & 0.210 \\
Sulphur, S & 16 & 29.8 & 0.019 & 0.00025 \\
Calcium, Ca & 20 & 37.2 & 0.0 & 0.0253 \\
Iron, Fe & 26 & 52.1 & 0.855 & 0.0626 \\
Nickel, Ni & 28 & 58.7 & 0.052 & 0.00196 \\
\hline \hline
\end{tabularx}
\caption{{Earth's elements including their atomic number and  mass. The last two columns show the mass fraction of each element in the Earth’s core and mantle taken from \cite{EarthComposition}.}}
\label{EarthModel}
\end{center}
\end{table}

In the next section, we  calculate the ionization form factor for all the elements in Table\,\ref{EarthModel}. Combining the result with the absolute abundance of elements at arbitrary radius $r$ as demonstrated in Fig.\,\ref{nCM}, one can fully determine the scattering behavior of the BDP propagation in the Earth. 

\section{Dark Matter Induced Ionization}
\begin{figure*}[htbp] 
  	\centering
  	\includegraphics[width=15.cm,trim={0mm -20mm 0 0},clip]{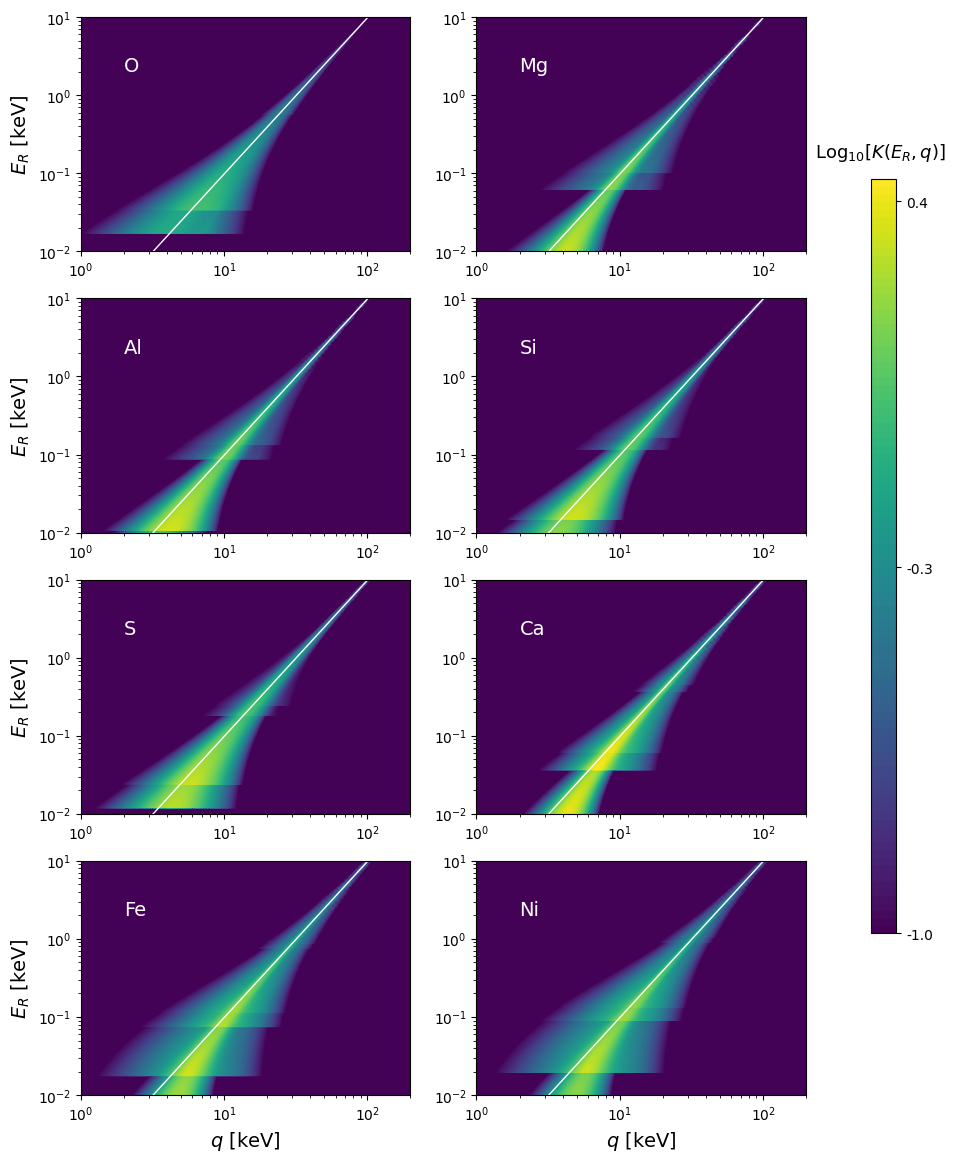}
  	    \vspace{-15mm}
  	\caption{
  		The atomic ionization form factor $K(E_{R}, q)$ for different atoms listed in Table\,\ref{EarthModel}  in the tight binding limit. The radial wave functions are determined using the RHF ground state wave functions in Eq. (\ref{RHFR}) with the coefficients $C_{j \ell n}, Z_{j \ell}$, $n_{j \ell}^{\prime}$ and binding energies provided in \cite{Bunge}. The $q$ and $E_{R}$ distribution converges to $E_{R}={q^2}/{(2m_{e})}$ in the large recoil energy limit, which is labeled by the  white solid line. 
  		}
  	\label{ARO}
\end{figure*}
In this section we briefly summarize how BDPs  ionize electrons bound inside atoms; for a more detailed discussion, see Appendices \ref{AA}, \ref{AB}, and \ref{CESI}. 
The  differential cross section for the ionization caused by an incoming BDP  $\chi$ (with velocity $v_\chi$) is given by
\begin{equation}
\frac{{\rm d}\sigma_{\mathrm{ion}} }{{\rm d} E_R} (v_\chi, E_R) = \frac{\overline{\sigma_e} m_{e} a_{0}^{2}}{2 \mu^2 v_\chi^2} \int_{q_{-}}^{q_{+}} \!\! q\  \left|F (q)\right|^{2}\ \!K(E_R, q) \ {\rm d} q,
\label{DCSI}
\end{equation}
where 
$\mu$ 
is the reduced mass of the BDP-electron system, 
$a_0 = 1/(\alpha m_{e})$ 
is the Bohr radius,  $q_{\pm} = m_\chi v_\chi 
\pm \sqrt{m_\chi^{2} v_\chi^{2}-2 m_\chi E_{R}}$ is the range for momentum transfer $q$, and $F(q)$ is the BDP form 
factor, which for the model described by Eq.\,({\ref{111}}) is
$F(q) = 1$.

The atomic form factor $K(E_{R}, q)$ for ionization describes the probability
of obtaining a particular recoil energy of an ionized electron for a given 
momentum transfer $q$. We follow the calculation presented in \cite{Catena:2019gfa,DarkARC}.
 The wave functions of the electron initial states with quantum numbers ($n, \ell$) are taken to be the Roothan-Hartree-Fock (RHF) ground state wave functions whose radial part is described by a linear combination of Slater-type orbitals,
\begin{equation}
R_{n \ell}(r) = a_{0}^{-3 / 2} \sum_{j} C_{j \ell n} \frac{\left(2 Z_{j \ell}\right)^{n_{j \ell}^{\prime}+1 / 2}}{\sqrt{\big(2 n_{j \ell}^{\prime}\big) !}} 
 \left(\frac{r}{a_{0}}\right)^{n_{j \ell}^{\prime}-1} \exp \left(-Z_{j \ell} \frac{r}{a_{0}}\right) .\label{RHFR}
\end{equation}
The values of the parameters $C_{j \ell n}, Z_{j \ell}$, $n_{j \ell}^{\prime}$, as well as the binding energies for each element are provided in  \cite{Bunge}. 
The final state wave functions, which are asymptotically free spherical waves in a central potential, are  given in \cite{Bethe}. The atomic form factor
 $K(E_{R}, q)$ defined in \cite{Roberts:2016xfw,Roberts:2019chv} is related to the ionization response function $f_{\text {ion }}^{n \ell}\left(k^{\prime}, q\right)$ through 
\begin{equation} 
K(E_R, q) = \sum_{n\ell} 
\frac{\left|f_{\text {ion }}^{n \ell}\left(k^{\prime}, q\right)
\right|^{2}}{2 k^{\prime 2} a_0^2} \Theta (E_R + E_B^{n\ell}) \ ,
\end{equation} 
where $\Theta$ is the Heaviside function. We have $E_R = - E_B^{n\ell} + k^{\prime 2} /2 m_{e}$, where $E_B^{n\ell}$ is the binding energy of the initial state electron, 
and $k^\prime$ is the momentum of the final state ionized electron. 
We take into account contributions from all accessible  states. A detailed calculation of $K(E_{R}, q)$ is presented  in Appendices \ref{AA} and \ref{AB}.

For the energy regime of the  BDP scenario considered here, the energy losses are dominated by the ionization process.
Scattering with the valence and conducting electrons, due to their small binding energy, should recover the elastic scattering limit at $E_R \sim \mathcal{O} (1)$ keV.
We thus treat the electron ionization in the tight binding approximation, where the electrons are assumed to have limited interactions with the neighboring atoms,  
so that the uncertainty of the molecular composition can be  ignored. 
We also neglect the dissipation induced by the transitions of an electron among bound states  (see \cite{Kouvaris:2014lpa, Emken:2019tni}), since these are subdominant compared to the ionization when the typical recoil energy $E_R$ is much larger than the binding energy of valence electrons. 
Under those assumptions, we calculate the ionization form factor $K(E_{R}, q)$ for each of the elements listed in  Table\,\ref{EarthModel} and show the results  in Fig.\,\ref{ARO}. For all the cases, as expected, the ionization form factors approach the kinetic region of elastic scattering, i.e., $E_R = q^2/(2 m_e)$, when $E_R$ is much larger than the binding energy. On the other hand, when $E_R$ is just enough to ionize an electron, $q$ has a broader  distribution.

\section{Propagation of Boosted Dark Particles Inside the Earth}

\subsection{Overview of the Monte Carlo simulation}

\begin{figure}[t!]
    \centering
    \includegraphics[scale=0.4]{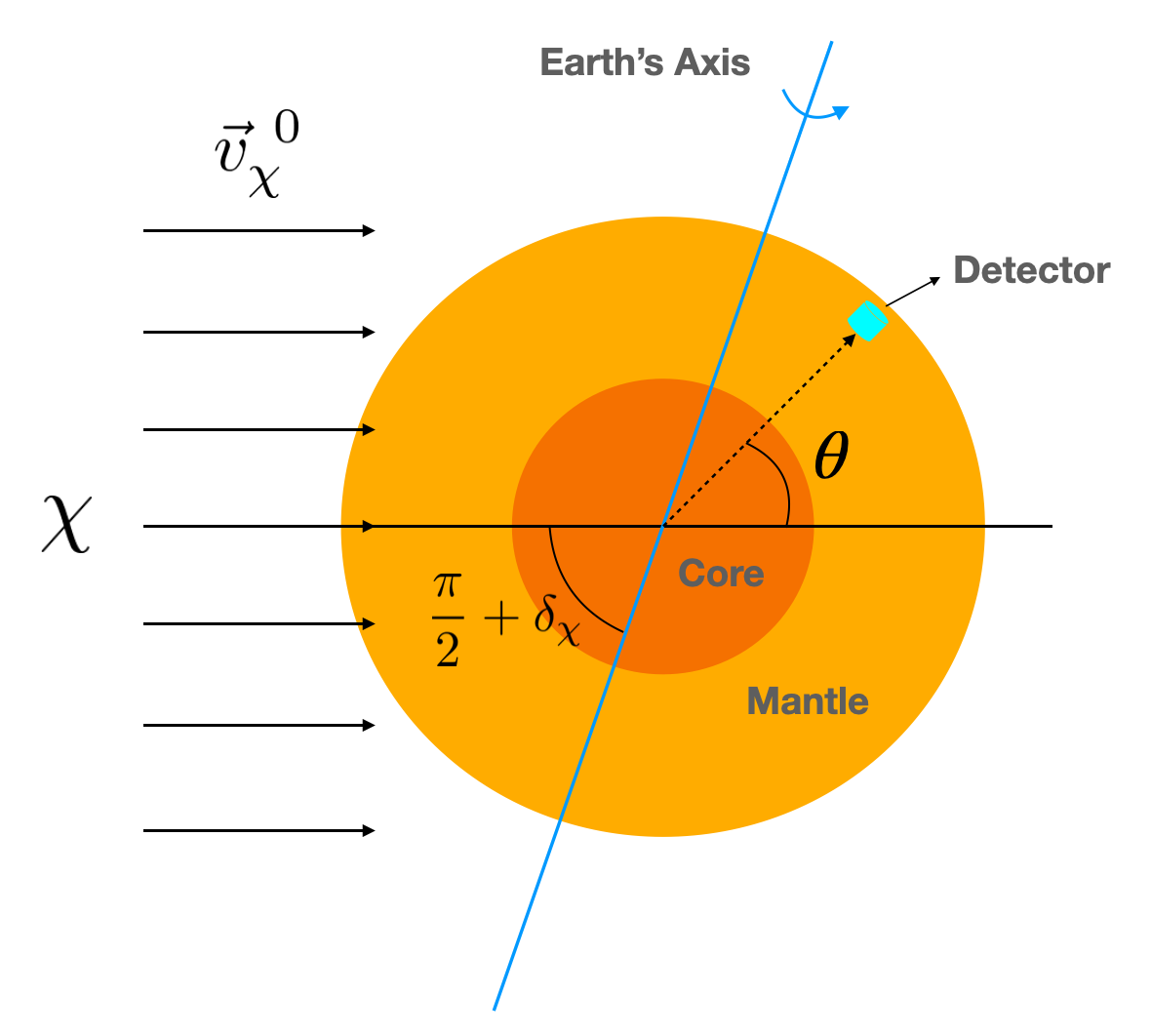}
    \caption{
        {Schematic plot showing a BDP flux arriving from a particular direction with velocity $\vec{v}_\chi^{\ 0}$. The polar angle $\theta$ is between the direction of the initial flux and the direction pointing from the Earth's center to the detector; the angle $\delta_\chi$ is the declination of the BDP flux direction in the equatorial coordinate system, ranging from $-\pi/2$ in the south to  $\pi/2$ in the north. 
        The blue and black solid lines denote the Earth's rotation axis and the incident direction of the flux, respectively. The light  and dark orange regions correspond to the Earth's mantle and core, and the cyan cylinder denotes the detector.}
    }
    \label{Theta}
\end{figure}

We assume that the BDPs are produced monochromatically and arrive at Earth from a fixed direction \cite{Fornal:2020npv}. 
Thus, the incoming BDP flux can be written as\footnote{
We confirmed that the effects of the BDP interactions  with the galactic medium and the Earth's atmosphere on its velocity distribution  are negligible compared to the effect of the interactions inside the Earth as it travels to the detector. Given how the latter affects the BDP velocity distribution, the former does not introduce a sizable modification to our assumption of a monochromatic energy spectrum for the BDP flux.
The influence of the atmosphere to hadrophilic dark matter was discussed, for example, in \cite{Bramante:2022pmn}.
}
\begin{equation} 
\frac{{\rm d}\Phi_{\rm init}}{{\rm d}^3 \vec{v}_\chi} = \Phi_0\ \delta^3 ( \vec{v}_\chi - \vec{v}_\chi^{\ 0}) \ ,\label{MIS}
\end{equation}
where $\Phi_0$ is the total initial flux directed towards the Earth. A schematic diagram of the model is shown in Fig.\,\ref{Theta}.

In order to understand the propagation of the BDP inside the Earth, one first needs to consider the interaction between the BDP and the  Earth's elements. According to Eq.\,(\ref{DCSI}), the probability distribution of the BDP final state after scattering is fully determined by the ionization form factor $K(E_R,q)$, where $E_R$ is the recoil energy and $q$ is the momentum transfer. From Eq.\,(\ref{DCSI}),  the mean free path of the BDP inside the Earth can be calculated as
\begin{equation}{\label{mfp}}
l_{\mathrm{fp}}^{\mathrm{ion}}(r,v_{\chi}, m_\chi) = \left[\sum_a n^a (r) \sigma_{\mathrm{ion}}^a(v_{\chi}, m_\chi) \right]^{-1},
\end{equation}
where the index $a$ denotes the type of the Earth's element. $n^{{a}}(r)$ is the number density of element $a$ at  radius $r$, which can be calculated from Fig.\,\ref{nCM}. $\sigma_{\mathrm{ion}}^{{a}}(v_{\chi}, m_\chi)$ is the ionization cross section between element $a$ and a BDP with velocity $v_{\chi}$ and mass $m_\chi$, obtained by integrating out the recoil energy $E_R$ and momentum transfer $q$ in Eq.\,(\ref{DCSI}).

\begin{figure}[t!]
    \centering
    \includegraphics[width=1\columnwidth]{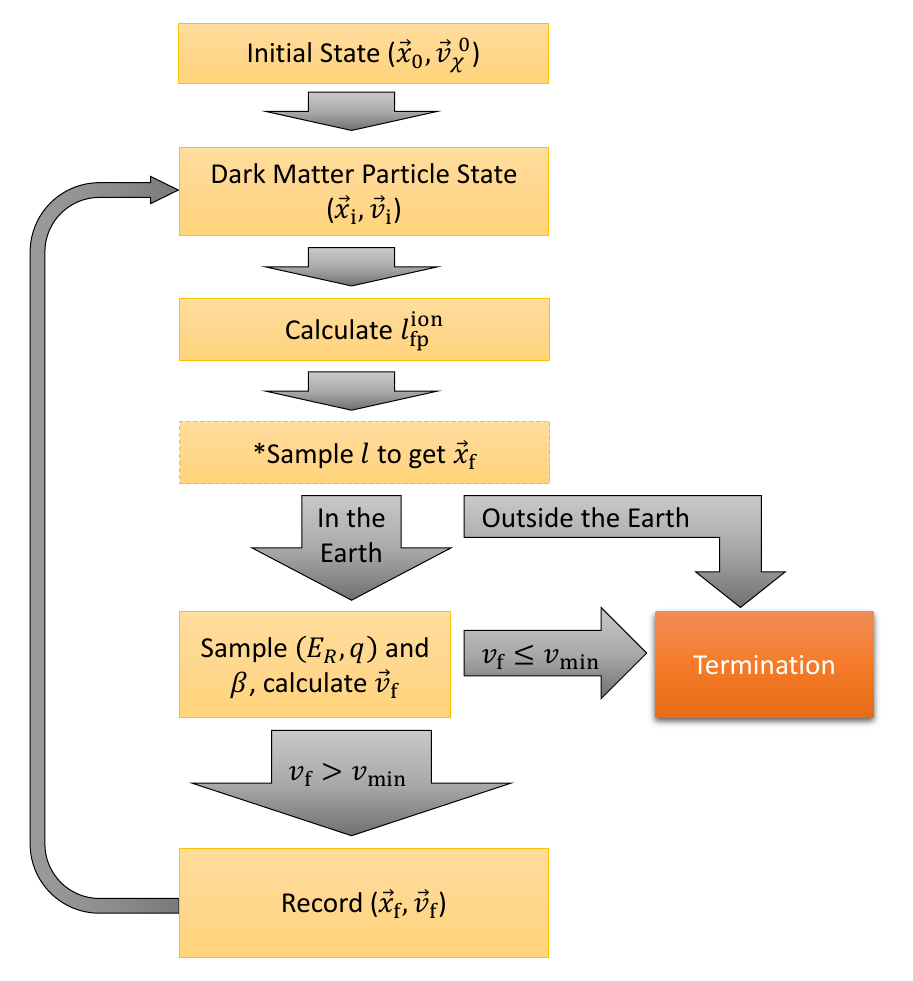}
    \caption{{The flow chart of the Monte Carlo simulation. The indices ${\rm i}$ and ${\rm f}$ are used to denote the initial and final states during each iteration.
    In each sampling of $l$, we check  if the path passes through the mantle-core border; if true, we use the coordinates where the path hits the mantle-border as the new starting point $\vec{x}_{\rm i}$, and we keep $\vec{v}_{\rm i}$ unchanged; this step is not shown in the flow chart. See Appendix \ref{sec_simulation}} for a more detailed discussion.}
    \label{simulation}
\end{figure}

We have developed a Monte Carlo simulation to study the BDP propagation inside the Earth. The flow chart of the simulation is shown in Fig.\,\ref{simulation}, and a  more detailed description can be found in Appendix \ref{sec_simulation}.
We start with BDPs  of mass $m_\chi$ and  velocity $\vec v^{\ 0}_\chi$, evenly distributed on the plane perpendicular to $\vec v^{\ 0}_\chi$. The main structure of the simulation is the iteration of scattering (${\rm i}$ and ${\rm f}$ denote the initial and final state for each step, respectively).  
In each iteration, we first calculate the mean free path $l_{\mathrm{fp}}^{\mathrm{ion}}(x_{\rm i}, v_{\rm i}, m_\chi)$ using Eq.\,(\ref{mfp}). Next, we use the exponential distribution $e^{-l/l_{\mathrm{fp}}^{\mathrm{ion}}} /l_{\mathrm{fp}}^{\mathrm{ion}}$ to sample $l$, which denotes the propagation distance for the BDP in this step of the iteration. 
The final position of the BDP, $\vec{x}_{\rm f}$ , can thus be easily calculated from $\vec{x}_{\rm i}$, $\vec{v}_{\rm i}$ and $l$.
Then, we sample the recoil energy $E_R$ and the momentum transfer $q$ whose probability distribution is proportional to $q\times K(E_R,q)$ according to the differential cross section in Eq.\,(\ref{DCSI}). Meanwhile the azimuthal angle $\beta$ on the transverse plane with respect to the initial velocity 
is drawn from a flat distribution between $0$ and $2\pi$. The values of $E_R$, $q$, and $\beta$ fully determine the momentum transfer vector $\vec{q}$, which is used to calculate the final velocity $\vec{v}_{\rm f}$.  
Lastly, the pair ($\vec{x}_{\rm f}$, $\vec{v}_{\rm f}$) is used as the input for the next iteration as ($\vec{x}_{\rm i}$, $\vec{v}_{\rm i}$).

Additionally, in each iteration we check whether the trajectory crosses the mantle-core border. If it does, we recalculate the mean free path $l_{\mathrm{fp}}^{\mathrm{ion}}$ and reset the starting point for this iteration at the spot where the crossing happens. Furthermore, the initial velocity $\vec{v}_{\rm i}$ remains unchanged.
The location and velocity at each iteration are recorded. The simulation stops once the BDP exits the Earth or when its velocity is smaller than the threshold velocity, which is either the  DM virial velocity or the minimum velocity to ionize an electron in xenon. For more details, please see Appendix \ref{PIE}. Finally, we perform the simulation with different BDP's initial velocity directions to account for the effect of Earth's rotation, as demonstrated in Fig.\,\ref{Theta}. 

\subsection{Distortion of the velocity distribution} 

Due to propagation inside the Earth, the BDP velocity distribution is distorted when reaching the detector. The amount of distortion depends on the polar angle $\theta$ between the incoming BDP flux and the direction pointing from the Earth's center to the detector, as shown in Fig.\,\ref{Theta}. Before showing the results of the Monte Carlo simulation, we first present a qualitative estimate of the distortion of the BDP velocity distribution.

The distance traveled $l$ inside the Earth depends on the depth of the detector $d$ and the direction of the incoming BDP flux. In terms of  $\theta$, it can be written as
\be
l = \sqrt{R_{E}^{2}-R_{D}^{2} \sin^2 \theta} + R_{D}\,\cos{\theta}
 \label{tdl}
\ee
where $R_{D} \equiv R_{E} -d$. In the limit $d \ll R_E$, $l$ ranges from $d$ to $\sqrt{2 R_{E} d}$ on the near side ($\frac{\pi}{2} < \theta \leq \pi$) and 
from $\sqrt{2 R_{E} d}$ to $2 R_{E}$ on the far side ($0\leq \theta < \frac{\pi}{2}$).
The BDP kinetic energy, $E_{\textrm{kin}} \equiv m_{\chi}v_{\chi}^2/2$, is smeared due to dissipation from ionization. For each scattering, the typical energy loss in the elastic scattering limit is $m_e v_\chi^2$ when the BDP is much heavier than electrons \cite{Fornal:2020npv} (see Appendix \ref{CESI} for a more detailed discussion). Thus the energy dissipation can be approximated in terms of the mean free path $l_{\mathrm{fp}}^{\mathrm{ion}}$ as 
\begin{equation}
    \frac{{\rm d} E_{\textrm{kin}}}{{\rm d} x} \simeq -\frac{m_{e}v_{\chi}^{2}}{l_{\mathrm{fp}}^{\mathrm{ion}}} \ , \label{dEkin}
\end{equation}
from which one can derive the dissipation of velocity  as
\begin{equation}
    v_{\chi} \left( l \right) \simeq v_{\chi} \left( 0 \right) \exp\left(- \int_0^l \frac{m_{e}}{m_{\chi}\,l_{\mathrm{fp}}^{\mathrm{ion}} (x)}\ {\rm d} x \right).\label{vfvi}
\end{equation}


In the elastic scattering approximation, the mean free path can be written as $l_{\mathrm{fp}}^{\mathrm{free}}\left( r\right) = \left[ n_e(r) \overline{\sigma_{e}} \right]^{-1}$, where $n_e(r) = \sum_a n^a (r) Z^a$ is the electron density including the contributions of all elements inside the Earth and $\overline{\sigma_{e}}$ is the scattering cross section between the BDP and a free electron. In Fig.\,\ref{CompMFP}, we compare the mean free path for ionization, $l_{\mathrm{fp}}^{\mathrm{ion}}$, with the one from elastic scattering, $l_{\mathrm{fp}}^{\mathrm{free}}$. At low BDP velocities, the finite binding energy suppresses the ionization. On the other hand, when $v_\chi \gg 10^{-2} \,c$, $l_{\mathrm{fp}}^{\mathrm{free}}(r)$ serves as a good approximation for $ l_{\mathrm{fp}}^{\mathrm{ion}}(r)$. In this approximation, taking the electron number density as $1\times10^{24}/\textrm{cm}^3$ near the Earth's surface,
$1.3\times10^{24}/\textrm{cm}^3$ at the mantle, and $3\times10^{24}/\textrm{cm}^3$ at the core, the mean free path of the BDP in each region is $l_{\mathrm{fp}}^{\mathrm{S}} \sim 100\, \textrm{m} \times \left( \mathrm{{10^{-28}\, cm^{2}}/{\overline{\sigma_{e}}}} \right)$, $l_{\mathrm{fp}}^{\mathrm{M}} \sim 75\, \textrm{m} \times \left( \mathrm{{10^{-28}\, cm^{2}}/{\overline{\sigma_{e}}}} \right)$, and $l_{\mathrm{fp}}^{\mathrm{C}} \sim 33\, \textrm{m} \times \left( \mathrm{{10^{-28}\, cm^{2}}/{\overline{\sigma_{e}}}} \right)$, respectively.


According to Eq.\,(\ref{vfvi}), one can define the effective distance at which the velocity distortion is significant,
\be l_{\textrm{eff}} \equiv l_{\mathrm{fp}}^{\mathrm{ion}} \frac{m_\chi}{m_e} \ .\ee This can be used to classify the distortion of the velocity distribution into  several cases:

\begin{figure}[t!]
    \centering
    \includegraphics[scale=0.60,clip]{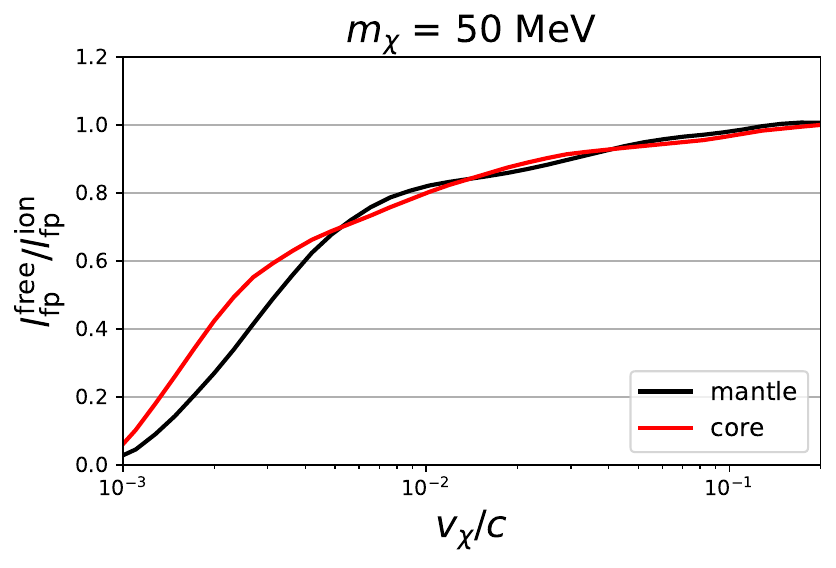}
    \caption{The ratio of the mean free path of the elastic scattering $l_{\textrm{fp}}^{\textrm{free}}$ and that of the ionization $l_{\textrm{fp}}^{\textrm{ion}}$, as a function of the BDP velocity. It converges to 1 for large $v_\chi$. We take  $m_\chi=50$ MeV as a benchmark.}
    \label{CompMFP}
\end{figure}

\begin{itemize}
\item[$\bullet$] $l_{\textrm{eff}} \ll d \simeq 1.6$  km ($\overline{\sigma_{e}} \gg 1 \times10^{-27} \,\mathrm{cm^{2}}$ for $m_\chi = 100$ MeV): extremely strong interaction. No events are expected in the detector;

\item[$\bullet$] $d$ $\ll l_{\textrm{eff}} \ll \sqrt{2 R_{E} d} \simeq 143$ km ($2 \times10^{-29} \ll \overline{\sigma_{e}} \ll 1 \times10^{-27} \,\mathrm{cm^{2}}$ for $m_\chi = 100$ MeV): strong interaction. No BDPs enter the detector if it is on the far side. The BDP velocity distribution may have a significant distortion when the BDP enters the detector on the near side;

\item[$\bullet$] $\sqrt{2 R_{E} d} \ll l_{\textrm{eff}} \ll 2 R_E \simeq 12740$ km ($1\times10^{-31} \ll \overline{\sigma_{e}} \ll 2\times10^{-29} \,\mathrm{cm^{2}}$ for $m_\chi = 100$ MeV): weak interaction. A significant distortion of the BDP velocity distribution may happen when the BDP enters the detector on the far side;

\item[$\bullet$] $l_{\textrm{eff}} \gg 2 R_E$ ($\overline{\sigma_{e}} \ll 1\times10^{-31} \,\mathrm{cm^{2}}$ for $m_\chi = 100$ MeV): extremely weak interaction. The BDP flux experiences almost no distortion of its velocity distribution.
\end{itemize}

\begin{figure*}[t!] 
  	\centering
  	\includegraphics[width=1.0\textwidth]{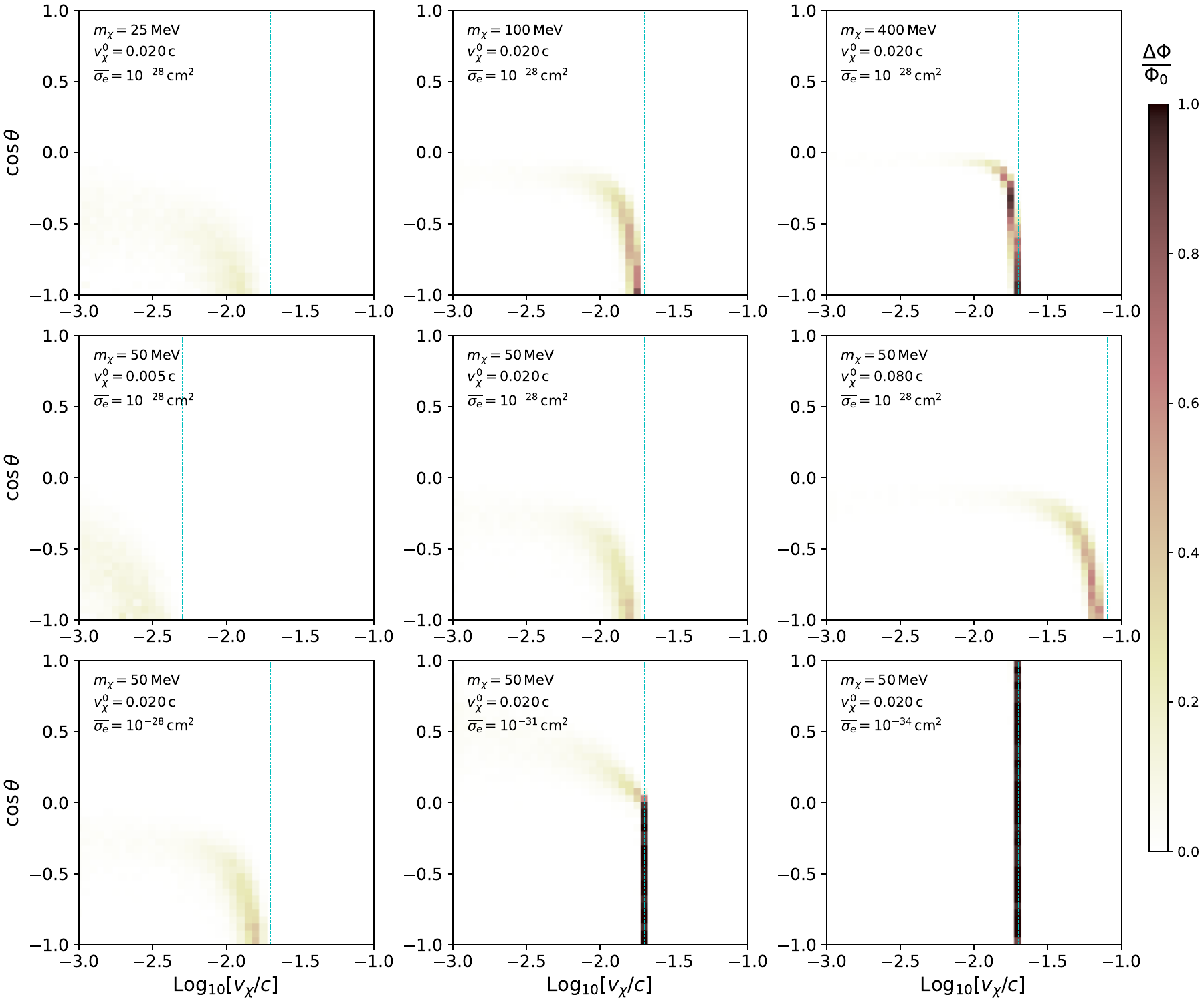}
  	\caption{
  	The BDP velocity distribution upon reaching the detector, after its propagation through the Earth. The three rows correspond to different choices of the BDP mass, the initial BDP velocity and the cross section, respectively. The value of the initial BDP velocity is indicated by the cyan dotted line. We use the color bar to characterize the normalized differential flux distribution as a function of the BDP velocity.  
  		$\Delta \Phi$ is the flux within each bin of ${\rm Log_{10}} [v_\chi/c]$.
  		}
  	\label{DT}
\end{figure*}  

In Fig.\,\ref{DT}, we show the results of our simulation for the BDP velocity distribution when it reaches the detector as a function of $\cos{\theta}$.
Nine cases are presented, illustrating how the differential velocity distribution depends on various model parameters.
The first row corresponds to different values of the BDP mass $m_\chi$.  Equation (\ref{vfvi}) implies that the larger the mass, the less distorted the velocity distribution is after scattering inside the Earth.  The second row corresponds to a variation of the initial velocity $v_\chi^0$. The third row compares three cases with various BDP-electron scattering cross sections $\overline{\sigma_e}$, corresponding to scenarios with strong interaction, weak interaction, and extremely weak interaction, respectively.

\section{Daily Modulation of Ionization Signals}
Due to Earth's rotation, the angle $\theta$ between the direction of the incoming BDP flux and the detector varies with a period of one day, 
\begin{equation}
\cos{\theta}(t)=-\cos{(\delta_{\chi})}\cos{(\delta_{D})}\cos{\left[ 2\pi \left( \frac{t-t_{0}}{24\textrm{h}} \right) \right]}
-\sin{(\delta_{\chi})}\sin{(\delta_{D})}\ ,
\end{equation}
where $\delta_\chi$ is the declination of the source of the BDP flux and $\delta_{D}$ is the detector's declination projected onto the celestial sphere. 
The time at which the BDP flux reaches at  upper culmination of the detector, denoted as $t_{0}$, can be determined using $2\pi t_{0}/(24{\rm h}) \equiv (\alpha_{\chi} - \lambda_{D})$ in terms of Greenwich Mean Sidereal Time (GMST). Here, $\alpha_{\chi}$ represents the right ascension of the BDP flux, and $\lambda_{D}$ represents the longitude of the detector. As an example, assuming a BDP flux from the GC with $\delta_{\chi,\textrm{GC}} \simeq -29.00^\circ$ and $\alpha_{\chi,\textrm{GC}} \simeq 266.40^\circ$, we find that the corresponding values of $t_0$ are $16.85$ h, $10.97$ h, and $0.63$ h for XENONnT ($\delta_{D} \simeq 42.47^\circ$, $\lambda_{D} \simeq 13.57^\circ$), PandaX ($\delta_{D} \simeq 28.20^\circ$, $\lambda_{D} \simeq 101.70^\circ$), and LUX-ZEPLIN ($\delta_{D} \simeq 44.35^\circ$, $\lambda_{D} \simeq -103.25^\circ$), respectively.
Since the GC is on the southern hemisphere and the three detectors we consider in this study are on the northern hemisphere, the detectors are on the far side of the Earth with respect to the BDP flux for the majority of the time.

\begin{figure}[t!]
    \centering
    \includegraphics[width = 1 \columnwidth]{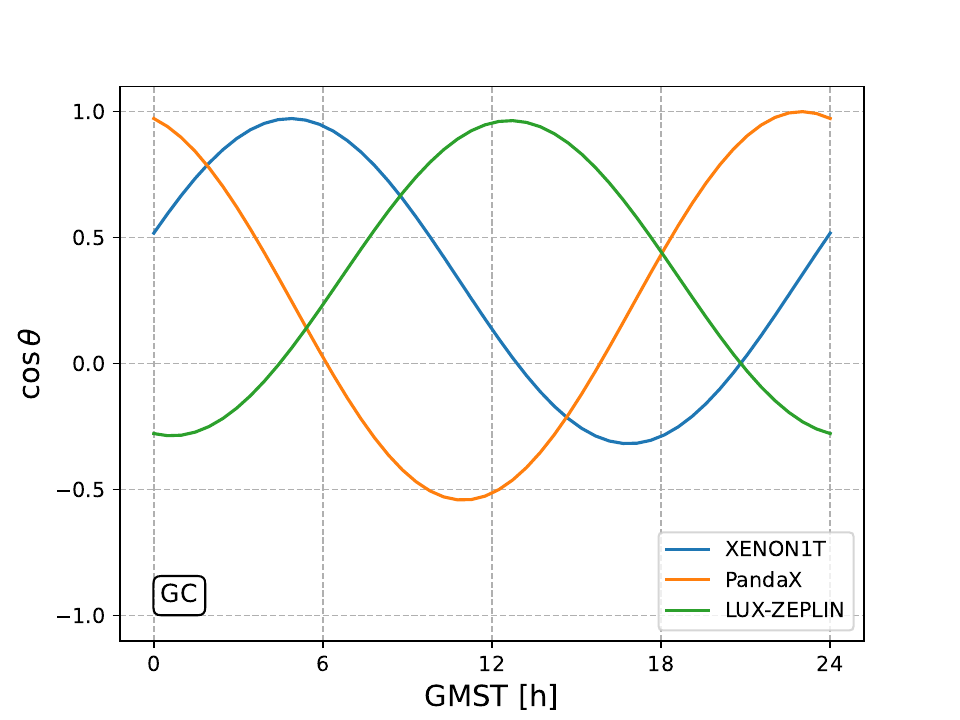}
     \includegraphics[width = 1 \columnwidth]{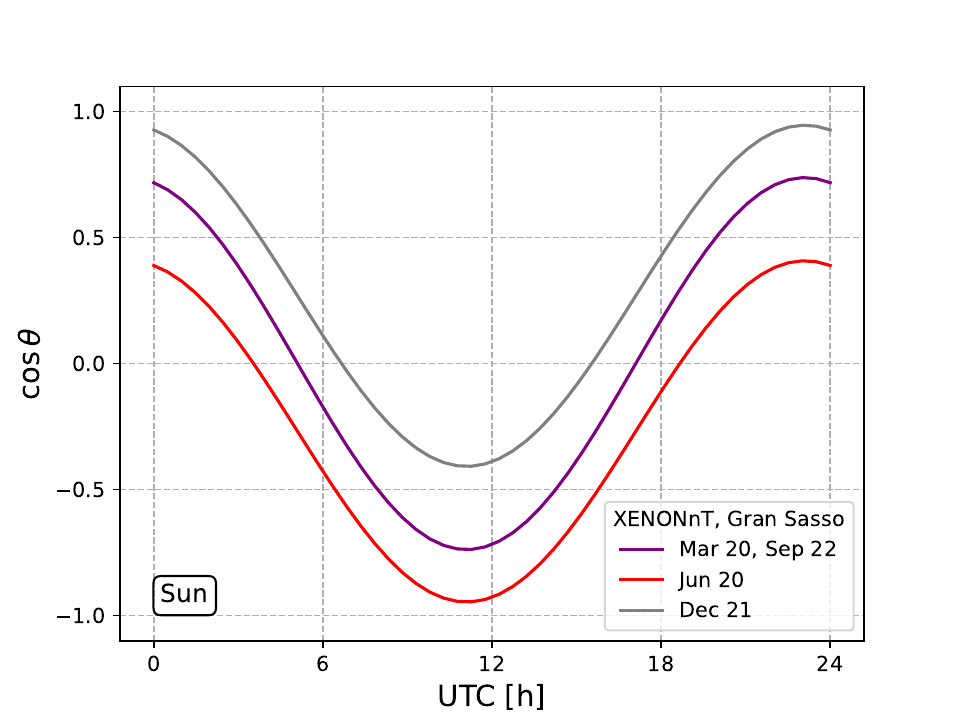}
    \caption{Top panel: Value of $\cos{\theta}$ as a function of the sidereal time for XENONnT, PandaX and LUX-ZEPLIN, respectively, assuming the BDP flux originates  in the GC. Bottom panel: The value of $\cos{\theta}$ as a function of UTC for the XENONnT experiment on four different days of the year, assuming the BDP flux arrives from the Sun.}
    \label{DD}
\end{figure}

Apart from BDP from the GC, one can also consider BDP from the Sun. In this case, the daily modulation is more conveniently described by the Coordinated Universal Time (UTC), shown in the bottom panel of Fig.\,\ref{DD}. The value of $\delta_{\chi,\textrm{Sun}} $ varies from $-23.5^\circ$ on December 21st to $23.5^\circ$ on June 20th. We take $t_0 = 11.1$ h according to the longitude of Gran Sasso.

The signal rate for each experiment can be written as
\begin{equation}
\frac{{\rm d}R}{{\rm d}E_R}=N_d \int\frac{{\rm d}\sigma_{\rm ion}}{{\rm d}E_R}
(v_{\chi},E_R)\,\frac{{\rm d}\Phi (v_\chi, \theta)}{{\rm d}v_{\chi}}\,
{\rm d}v_{\chi}\ ,\label{dRdER}
\end{equation}
where the differential cross section is provided in Eq.\,(\ref{DCSI}). $N_d\simeq4.2\times10^{27}$ ton$^{-1}$ is the number of xenon atoms in the detector. 
\begin{figure*}[htbp] 
  	\centering
  	\includegraphics[width=1.0 \textwidth]{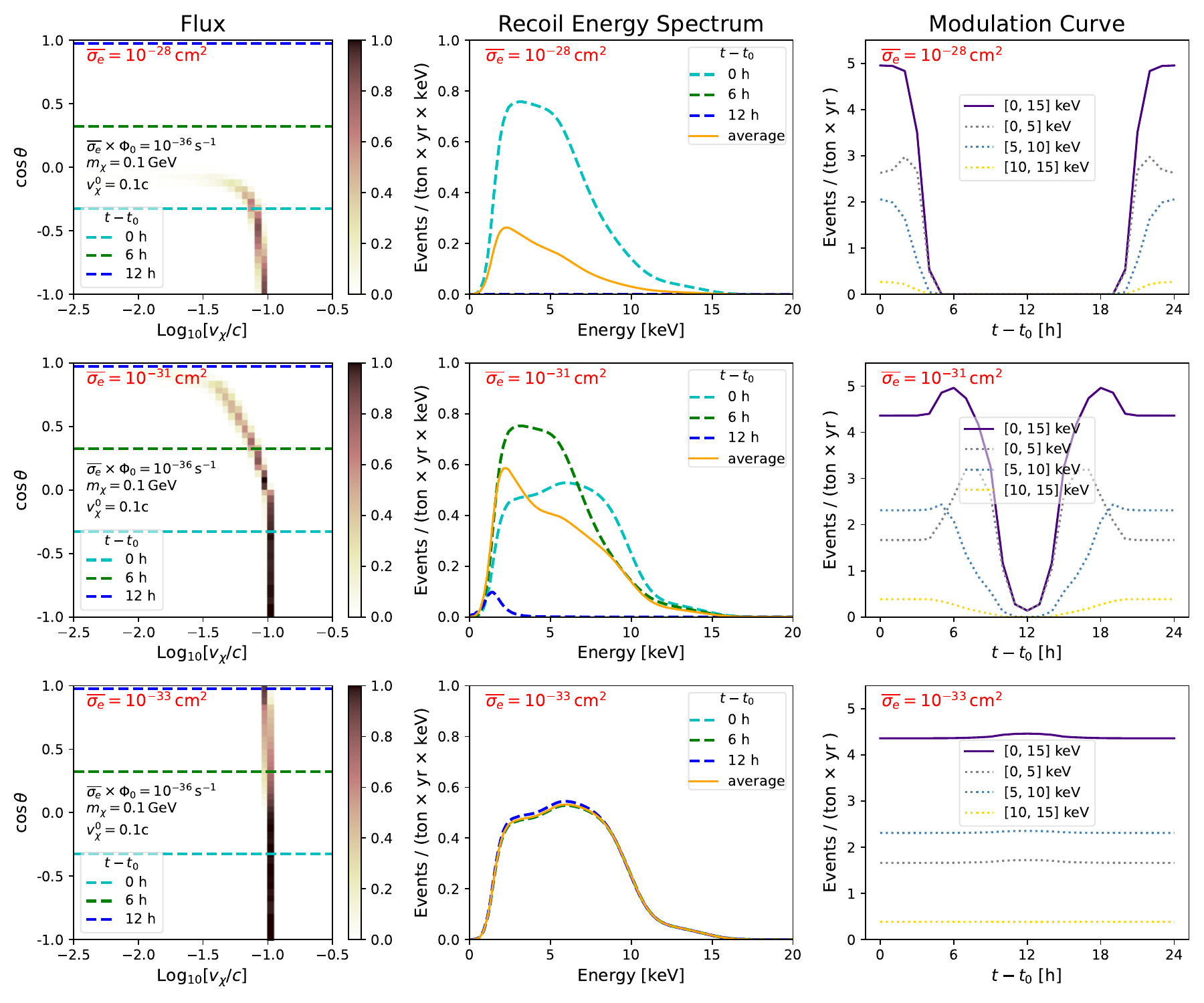}
  	\caption{The velocity distribution,   electron  recoil  energy  spectrum, and  time evolution of the event rate. The BDP flux is assumed to originate in the GC. The BDP mass $m_{\chi}$ and the initial velocity $v_{\chi}^{\,0}$ are taken to be $0.1\,\mathrm{GeV}$ and $0.1\,c$, respectively.  
  	The three benchmark values for the cross section considered are: $\overline{\sigma_{e}}=$  $10^{-28}\,\mathrm{cm^{2}}$ (upper panels),  $10^{-31}\,\mathrm{cm^{2}}$ (middle panels), and  $10^{-33}\,\mathrm{cm^{2}}$ (lower panels). In the left column, the velocity distribution as a function of $\cos \theta$ is presented for the three cases. 
  	To show the time dependence caused by the Earth shielding effect, the results at different sidereal times $(t - t_0) =0$ h, $6$ h, and $12$ h are highlighted by the cyan, green, and blue dashed lines, respectively, where $t_0$ corresponds to the time when the GC culminates over the  Gran Sasso detector. 
  		The middle column, with fixed $\overline{\sigma_e} \times \Phi_0 = 10^{-36} {\rm s}^{-1}$, shows the corresponding electron recoil energy spectrum for the three benchmark points at the three times. 
  The averaged signals are denoted by the orange solid lines.
  		In the right column, the time evolution of the signals for different recoil energy bins is presented.
  		}
  	\label{DR}
\end{figure*}  
The electron recoil energy spectrum varies with time.  

In Fig.\,\ref{DR}, we present our results for the BDP flux from the GC for $m_\chi = 100$ MeV, $v_\chi^0 = 0.1 c$, and  three benchmark cross sections: $\overline{\sigma_{e}}=$  $10^{-28}\,\mathrm{cm^{2}}$ (upper panels),  $10^{-31}\,\mathrm{cm^{2}}$ (middle panels), and  $10^{-33}\,\mathrm{cm^{2}}$ (lower panels). The product of the total flux $\Phi_0$ and $\overline{\sigma_e}$ is fixed to be $\overline{\sigma_e} \times \Phi_0 = 10^{-36} {\rm s}^{-1}$, as discussed in Appendix\,\ref{constraints}, to satisfy the XENONnT constraint \cite{XENONCollaboration:2022kmb}.

The first column of Fig.\,\ref{DR} shows the velocity distribution for the three cases. As expected, the distortion of the flux varies when changing the ratio between $\Phi_0$ and $\overline{\sigma_e}$. The three dashed lines correspond to the values of $\cos\theta$ for the detector at Gran Sasso at $t - t_0 =$  $0$ h, $6$ h, and $12$ h. In the most distorted case, i.e., when $\overline{\sigma_{e}}=$  $10^{-28}\,\mathrm{cm^{2}}$, the flux is completely shielded at  $t - t_0 = 12$ h, while for $\overline{\sigma_{e}}=$  $10^{-33}\,\mathrm{cm^{2}}$ the distortion of the flux is negligible. The second column shows the recoil energy spectrum from Eq.\,(\ref{dRdER}) at the three different time $t-t_0$. The distortion of the flux leads to a shift of the recoil energy events towards  lower energy bins. In addition, a time-averaged spectrum is denoted by the orange solid line. Finally, the last column shows the time evolution of the event rate in the three bins: [0, 5 keV], [5 keV, 10 keV], [10 keV, 15 keV], and the sum of those three bins.

All those unique features in the experimental data can be used to extract the properties of the BDP in a systematic and comprehensive manner.


One way to examine the daily modulation signal is to perform a Fourier transform on the data \cite{Lee:2015qva}. The signal can be parametrized as
\begin{equation}
\frac{{\rm d} R}{{\rm d} E_R}=A_{0}+\sum_{n=1}^{\infty}\left[A_{n} \cos \left(  2\pi n  \frac{t-t_{n}}{T}\right)\right],
\end{equation}
where $T$ is the modulation period. For example, $T$ is either one sidereal day or one solar day (when the BDP source is the GC or the Sun, respectively). $A_n$ is the amplitude in the Fourier series and $t_n$ is the relative phase.
For the signal, $t_n$  converges to $t_0$   for a given recoil energy.
%
A fit to  $t_n$ provides  information on the direction of the BDP flux.
If we correlate the time series of the signals for three different detectors, one expects the differences in $t_0$ for each detector to be related to the differences in the detector locations.

\section{Discussion}

In this paper, we carried out a detailed analysis of the  daily modulation of the signal expected from BDPs interacting with electrons. Such an effect can be searched for in  terrestrial DM direct detection experiments such as XENONnT, PandaX and LUX-ZEPLIN. 

In particular, we developed  a Monte Carlo code to simulate the BDP's trajectory through the Earth to the detector. 
Considering a benchmark scenario with the BDP source located at the GC, we calculated the expected time variation of the signal in terms of the electron recoil event rate. Our predictions can be directly compared to current and future data.

It is worth noting that a different study regarding the daily modulation of the BDP signal has been carried out in \cite{Ge:2020yuf}, where the BDP is being considered to be produced by cosmic ray scattering. Instead of the BDP-electron scattering, the focus of that work is on the hadronic interaction of the BDP.
The hadronic form factor suppresses the interaction probability at large momentum transfer, in which case the distortion of the flux becomes most pronounced in the intermediate energy regime. Combined with the difference in the initial velocity distribution of the BDP flux, this leads to substantially different predictions for the BDP energy spectrum in the detector after the signal propagated through  Earth.

A possible future extension of our work includes  calculating 
local geographic effects on the BDP signal. It would also be interesting to consider the energy loss through the excitation among various atomic bound states, which would require a more in-depth knowledge of the chemical composition of the Earth.

\section*{Acknowledgements} 
We are grateful to  Ran Ding, Timon Emken, Shao-Feng Ge, Benjamin Roberts, Ningqiang Song, and Yanjie Zeng for useful discussions. 
Y.C. is supported by Villum Investigator program supported by the VILLUM Foundation (Grant No. VIL37766) and the DNRF Chair program (Grant No. DNRF162) by the Danish National Research Foundation, and by FCT (Fundação para a Ciência e Tecnologia I.P, Portugal) under Project No. 2022.01324.PTDC, and by the China Postdoctoral Science Foundation under Grants No. 2020T130661, No. 2020M680688, the International Postdoctoral Exchange Fellowship Program, and by the National Natural Science Foundation of China (NSFC) under Grant No. 12047557. 
J.S. is supported by the National Natural Science Foundation of China under Grants No. 12025507, No. 11690022, No.11947302; and is supported by the Strategic Priority Research Program and Key Research Program of Frontier Science of the Chinese Academy of Sciences under Grants No. XDB21010200, No. XDB23010000, and No. ZDBS-LY-7003 and CAS project for Young Scientists in Basic Research YSBR-006.
P.S. is supported by NSF Grabt PHY-2014075. 
X.X. is supported by Deutsche Forschungsgemeinschaft under Germany’s Excellence Strategy EXC2121 “Quantum Universe” - 390833306.
Y.Z. is supported by U.S. Department of Energy under Award No. DESC0009959.
Y.Z. would like to thank the ITP-CAS for their kind hospitality.

\appendix

\section{From elastic scattering to ionization}\label{AA}
The differential cross section for the $2 \to 2$ elastic BDP-electron scattering  is,
\ba
{\rm d} \sigma_{\mathrm{free}}=\frac{\overline{\left|\mathcal{M}_{\mathrm{free}} \right|^{2}}}{4 E_{\chi} E_{e} v_\chi}   (2 \pi)^{4} \delta^{4}\left(k+p-k^{\prime}-p^{\prime}\right) \times
\\ \frac{{\rm d}^{3} p^{\prime}}{(2 \pi)^{3}\ 2 
E_{\chi}^{\prime}} \frac{{\rm d}^{3} k^{\prime}}{(2 \pi)^{3}\ 2 E_{e}^{\prime}} \ , \label{dcf}
\ea
where $p$ and $k$ are the four-momenta of the BDP and electron initial state, while the prime denotes the final state. The $v_\chi$ is the relative velocity between the BDP and electron initial states. The \(\mathcal{M}_{\mathrm{free}}\) is the matrix element for the elastic scattering which depends on the momentum transfer $\vec{q} \equiv \vec{p} - \vec{p}^{\,\prime}$ in the nonrelativistic limit of electrons. 

In the elastic scattering, the initial and final state electron wave functions are taken to be plane waves $|  e_{\vec{k}} \rangle$. However, for a process like ionization, both the initial bound state and the final unbound state are represented by a wave packet in the momentum space,
\begin{equation}
|  e_{\vec{k}} \rangle \longrightarrow \int \frac{\sqrt{V} {\rm d}^{3} k}{(2 \pi)^{3}} {\psi}_{i}(\vec{k}) |e_{\vec{k}} \rangle, 
\end{equation}
where $V \equiv (2\pi)^3 \delta^3 (\vec{0})$ is the volume of space. The momentum space wave functions  satisfy the normalization condition $\int d^{3} k  | {\psi}(\vec{k}) |^2 / (2 \pi)^{3} = 1$. In the nonrelativistic limit, the scattering amplitude becomes
\bea  
&&(2 \pi)^{3} \delta^{3}\left(\vec{k} + \vec{q}-\vec{k}^{\prime}\right) \mathcal{M}_{\mathrm{free}} (\vec{q}) \\
&\rightarrow& \int \frac{V {\rm d}^{3} k}{(2 \pi)^{3}} {\psi}_{f}^{*}(\vec{k}+\vec{q}) {\psi}_{i}(\vec{k})\ \mathcal{M}_{\mathrm{free}} (\vec{q})\nn\\
&=& V f_{i \rightarrow f} (\vec{q})\ \mathcal{M}_{\mathrm{free}} (\vec{q}), \label{SAb}
\eea
where we define the transition form factor as
\bea f_{i \rightarrow f} (\vec{q}) &\equiv& \int \frac{{\rm d}^{3} k}{(2 \pi)^{3}} {\psi}_{f}^{*}(\vec{k}+\vec{q}) {\psi}_{i}(\vec{k})\\
&=& \int {\rm d}^{3} x\ \psi_{f}^{*}(\vec{x}) e^{i \vec{x} \cdot \vec{q}} \psi_{i}(\vec{x}). \label{tff}\eea
It describes the transition probability with a given momentum transfer $\vec{q}$. We can further rewrite the form factor in terms of the coordinate space wave functions, which becomes $\psi(\vec{x})=\exp (i \vec{k} \cdot \vec{x}) / \sqrt{V}$ in the plane wave limit.

For the ionization process, one needs to perform the summation on all the bound electrons in the initial state, as well as the final state phase space. Using the quantum numbers ($n, \ell, m$) to label the initial bound state, one has
\be \sum_{\textrm{occupied}} = g_s \sum_{n, \ell, m},\label{iPS}\ee
where $g_s = 1$ or $2$ represents the occupancy due to the spin degeneracy.  The final states are characterized by the asymptotically free spherical wave functions with phase space summation written as
\be  \frac{{\rm d}^{3} k^{\prime}}{(2 \pi)^{3}}  \rightarrow \sum_{\ell^{\prime} m^{\prime}}  \frac{k^{\prime 2} {\rm d} k^{\prime}}{(2 \pi)^{3}} 
= \sum_{\ell^{\prime} m^{\prime}}  \frac{k^{\prime} m_e {\rm d} E_{R}}{(2 \pi)^{3}}\
 \Theta (E_R + E_B^{n\ell}), \label{fPS}\ee
where $\Theta$ is the Heaviside function. $E_B^{n\ell} < 0$ is the binding energy for a given initial bound state $(n, \ell)$. The recoil energy $E_R$ is the sum of the final state electron kinetic energy and the amount of binding energy
$ E_R = k^{\prime 2}/2m_e + |E_B^{n\ell}|$.

Finally, one has to perform the integral over the momentum of the BDP final state. With  ${\rm d}^3 p^\prime = {\rm d}^3 q$, the energy conservation leads to
\be \frac{{\rm d}^3 q}{(2\pi)^2}\ \delta \left( E_R + \frac{q^2}{2 m_\chi} - q v_\chi \cos \theta_{qv_\chi} \right) 
= \frac{q {\rm d} q}{2\pi v_\chi}\label{dq}\ee
with integration limits 
\be q_{\pm} = m_{\chi} v_\chi \pm \sqrt{m_{\chi}^2 v_\chi^2 - 2 m_\chi E_R} \ .\ee

Putting everything together, in the nonrelativistic limit, the ionization differential cross section for the BDP with velocity $v_\chi$ can be written as
\be
\frac{{\rm d} {\sigma}_{\mathrm{ion}}}{{\rm d} E_R} = \sum_{n\ell}  \frac{ \overline{\sigma_{e}}\ \Theta (E_R + E_B^{n\ell})}{8 \mu^{2} v_\chi^2  (E_R + E_B^{n\ell})}    \int^{q_+}_{q_-}  q  \left| F\left( q\right)\right|^{2} \left|f_{\text {ion }}^{n \ell}\left(k^{\prime}, q\right)\right|^{2} {\rm d} q,\label{difCSion}
\ee
where $\overline{\sigma_{e}} \equiv \mu^{2}
\overline{\left|\mathcal{M}_{\text {free }}\left(\alpha m_{e}\right)
\right|^{2}} / \left(16 \pi m_{\chi}^{2} m_{\rm e}^{2}\right)$ 
is the BDP-electron elastic scattering cross section evaluated at $q = \alpha m_{e}$, and $\mu$ is the reduced mass. The function $F\left( q\right)$ is the BDP form factor,
and we have $F=1$ for the benchmark model studied here. The transition form factor defined in Eq. (\ref{tff}) for the ionization process can be written as
\be
\left|f_{\text {ion }}^{n\ell} \left(k^{\prime}, q\right)\right|^{2}=\frac{4 k^{\prime 3} V}{(2 \pi)^{3}} \sum_{\ell^{\prime} m^{\prime} m}\left|\int {\rm d}^{3} x \tilde{\psi}_{k^{\prime} \ell^{\prime} m^{\prime}}^{*}(\vec{x}) \psi_{i}^{n\ell m} (\vec{x}) e^{i \vec{q} \cdot \vec{x}}\right|^{2}.\label{fion}
\ee
It has no dependence on the direction of $\vec{q}$ due to the spherical symmetry. As mentioned in the main text, this is related to the factor $K (E_R, q)$ defined in \cite{Roberts:2016xfw,Roberts:2019chv} as 
\be K(E_R, q) = \sum_{n\ell} 
\frac{\left|f_{\text {ion }}^{n \ell}\left(k^{\prime}, q\right)
\right|^{2}}{2 k^{\prime 2} a_0^2} \Theta (E_R + E_B^{n\ell}),\ee 
where $a_0 = 1/(\alpha m_{e})$ 
is the Bohr radius.

\section{Atomic ionization factor}\label{AB}
The ionization form factor in Eq. (\ref{fion}) is obtained by calculating the spatial overlap between the initial and final electron wave functions, convoluted with the momentum transfer $e^{i \vec{q} \cdot \vec{x}}$. Following \cite{Essig:2012yx, Catena:2019gfa}, we expand  $e^{i \vec{q} \cdot \vec{x}}$ as a linear combination of spherical harmonic functions, and the Eq. (\ref{fion}) can be written as
\begin{widetext}
\ba  \left|f_{\text {ion }}^{n\ell} \left(k^{\prime}, q\right)\right|^{2} =
\frac{4 k^{\prime 3} V}{(2 \pi)^{3}} \sum_{\ell^{\prime}=0}^{\infty} \sum_{L=\left|\ell-\ell^{\prime}\right|}^{\ell+\ell^{\prime}}(2 \ell+1)\left(2 \ell^{\prime}+1\right) 
  (2 L+1)\left[\begin{array}{lll}
\ell & \ell^{\prime} & L \\
0 & 0 & 0
\end{array}\right]^{2} \left|I_{R}(q)\right|^{2}. \ea
\end{widetext}
Here $[\cdots]$ is the Wigner 3-$j$ symbol, and $I_{R}(q)$ is defined to be
\be I_{R}(q) \equiv \int \mathrm{d} r r^{2} R_{k^{\prime} \ell^{\prime}}^{*}(r) R_{n \ell}(r) j_{L}(q r),\label{IRQ}\ee
in which $j_{L}(q r)$ is the first order spherical Bessel function.
The radial wave functions of the electron initial bound state can be written as a sum of Slater-type orbital wave functions 
\be
R_{n \ell}(r) = a_{0}^{-3 / 2} \sum_{j} C_{j \ell n} \frac{\left(2 Z_{j \ell}\right)^{n_{j \ell}^{\prime}+1 / 2}}{\sqrt{\left(2 n_{j \ell}^{\prime}\right) !}} 
 \left(\frac{r}{a_{0}}\right)^{n_{j \ell}^{\prime}-1} \exp \left(-Z_{j \ell} \frac{r}{a_{0}}\right) ,
\ee
where the parameters $C_{j \ell n}, Z_{j \ell}$ and $n_{j \ell}^{\prime}$, for each atom species, are given in \cite{Bunge}.

The wave function of the ionized electron in the final state is the unbound solution of the Schrodinger equation with a hydrogen-like potential $-Z^{n\ell}_{\textrm{eff}}/r$. The numerical results are provided in \cite{Catena:2019gfa}, for example. This recovers the free plane wave solution when the kinetic energy is much larger than the binding energy. The effective charge, $Z^{n\ell}_{\textrm{eff}}$, is related to the binding energy as $Z^{n\ell}_{\textrm{eff}} = n \sqrt{-E_B^{n\ell} / 13.6\, \textrm{eV}}$ \cite{Bunge}. 

\section{Comparison between elastic scattering and ionization} \label{CESI}

Let us first consider the elastic scattering, assuming the electron is free and at rest. For a contact interaction with $F(q) = 1$, in the nonrelativistic approximation of the final state electron, one can derive the differential cross section as a function of the recoil energy, $E_R = k^{\prime 2}/2m_e$, as
\begin{equation} \frac{{\rm d} \sigma_{\mathrm{free}}}{{\rm d} E_{R}} = \frac{\overline{\sigma_e} m_e}{2 \mu^2 v_\chi^2}   \Theta\left( \frac{2 \mu^2 v_\chi^2}{m_e} - E_{R}\right),  \label{dcfdER} 
\end{equation}
which is a flat distribution for $E_R < 2 \mu^2 v_\chi^2/m_e$.

We now include the effect of the binding energy and consider the ionization process. This requires the velocity of the incoming BDP particle and its mass to be large so that the momentum transfer is enough to trigger the ionization. In this limit, the integration range ($q_-, q_+$) in Eq.\,(\ref{difCSion}) is ($E_R/v_\chi$, $2 m_\chi v_\chi$) at leading order, which effectively becomes ($0$, $\infty$) for the integral.
Thus below the energy cutoff in Eq. (\ref{dcfdER}), the ratio between the differential cross section for the ionization of an electron with ($n, \ell$), i.e. Eq.\,(\ref{difCSion}), and that of a free electron scattering, i.e. Eq. (\ref{dcfdER}), can be written as
\be n^{n \ell}_{\textrm{eff}} (E_R) \equiv \int^{\infty}_{0} q\ \left|f_{\text {ion }}^{n \ell}\left(E_R, q\right)\right|^{2} \frac{\Theta (E_R + E_B^{n\ell})}{4 m_e (E_R + E_B^{n\ell})}\ {\rm d} q.\label{eqneff}\ee
This is defined to be the effective electron number for a given atomic level. 
\begin{figure}[t!]
  	\centering
  	\includegraphics[scale=0.6,clip]{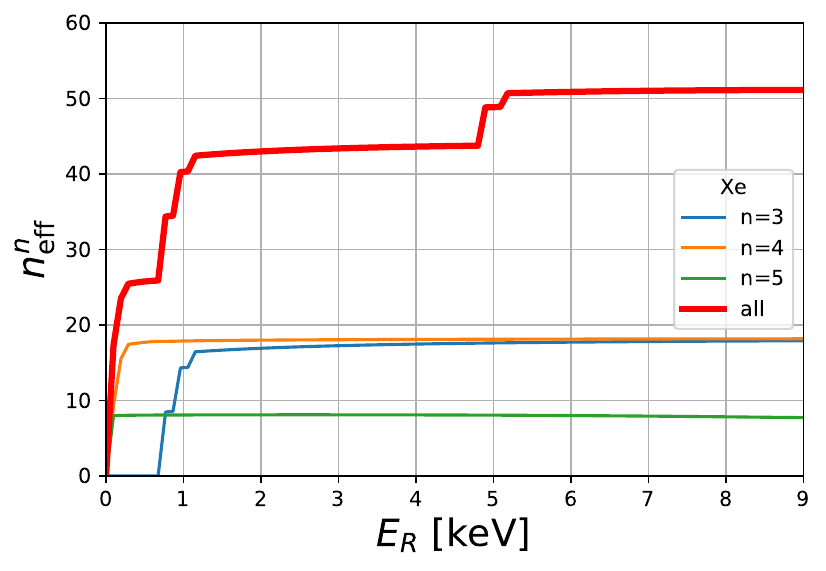}
  	\caption{The effective electron number for various xenon shells  in the large velocity and heavy BDP limit. In particular,  $n^{n}_{\textrm{eff}} \equiv \sum_\ell n^{n \ell}_{\textrm{eff}}$, as defined in Eq. (\ref{eqneff}).}
  	\label{neff}
  \end{figure}
The numerical results for xenon are shown in Fig.\,\ref{neff}. The results converge to the number of the electrons for that  shell. Notice that it requires the sum of the final state angular momentum number $\ell^\prime$ to a large number to properly mimic the final state wave function when $E_R$ is large. 

\begin{figure}[t!]
  	\centering
  	\includegraphics[width=1\columnwidth]{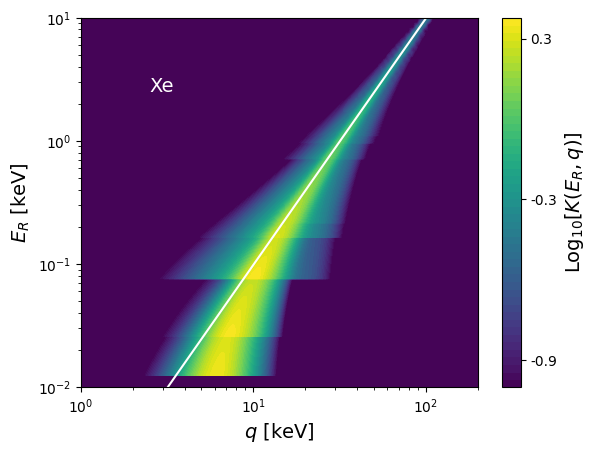}
  	\caption{The atomic ionization form factor $K(E_{R}, q)$ for xenon. The white solid line corresponds to $E_{R}={q^2}/{(2m_{\mathrm{e}})}$. 
  		}
  	\label{KdisERq}
  \end{figure}


This result is also consistent with the kinetic distribution of the ionization form factor in the large recoil energy limit. In Fig. \ref{KdisERq}, we show the result for xenon. In the limit of a large recoil energy, the form factor peaks at $E_R \simeq q^2/(2 m_e)$, which recovers the kinetic distribution of elastic scattering.


\section{Monte Carlo simulation}\label{sec_simulation}


\subsection{Initial setup}\label{IS}

The simulation starts with four input parameters: $m_{\chi}$, $\overline{\sigma_{e}}$, $v_{\chi}^{\,0}$, and $N$, which correspond to the BDP mass, the BDP-electron elastic scattering cross section evaluated at $q=\alpha m_{e}$, the initial incident BDP velocity, and the number of simulation events, respectively. We consider a BDP flux from the GC or the Sun. 
We set the direction of the $z$-axis to always coincide with the direction of the incoming BDP flux, see Fig.\,\ref{Theta} for an illustration.

To generate the initial BDP flux evenly distributed on the plane orthogonal to the $z$-direction, we first draw a random value from the uniform distribution $[0,R_{E}^2)$ where $R_E$ is the Earth's radius, then we define $\rho_{xy}$ as the square root of the previously generated number. Next, we draw a random azimuthal angle $\phi_{E}$ from a uniform distribution  $[0,2\pi)$. With this choice, we can calculate the position of each BDP particle entering the Earth in the Cartesian coordinate ($x, y, z$) as,
\begin{eqnarray}
x_{0} &=& \rho_{xy}  \cos{\phi_{E}} \ ,\nonumber \\
y_{0} &=& \rho_{xy}  \sin{\phi_{E}}\ ,\nonumber \\
z_{0} &=& \sqrt{R_{E}^2-x_{0}^2-y_{0}^2}\ .
\end{eqnarray}

\subsection{Propagation inside the earth}\label{PIE}

Next we consider the propagation of the BDP inside the Earth. 
The simulation procedure of the BDP propagation is shown on the flow chart in Fig.\,\ref{simulation}.
For each iteration, we first calculate the mean free path, $l_{\mathrm{fp}}^{\mathrm{ion}}$, of the BDP particle traveling inside the Earth, using Eq.\,(\ref{mfp}). Next, we determine the travel distance between two successive scatterings in the Earth's mantle or core according to an exponential probability distribution, 
\begin{equation}\label{mfp_app}
f\bigg(l;\frac{1}{l_{\mathrm{fp}}^{\mathrm{ion}}}\bigg) = \frac{1}{l_{\mathrm{fp}}^{\mathrm{ion}}}\exp \bigg(-\frac{l}{l_{\mathrm{fp}}^{\mathrm{ion}}}\bigg),
\end{equation}
Combining $l$  with the final velocity calculated from the previous step,  we obtain the endpoint of the trajectory in each iteration where the scattering occurs. It becomes subtle when the trajectory hits the mantle-core border before it ends. If this happens, we set the location where the trajectory hits the mantle-core border as the new starting point $\vec{x}_{\rm i}$ of this iteration while the velocity is left unchanged.

\begin{figure*}[t!] 
  	\centering
  	\includegraphics[width=0.8\textwidth]{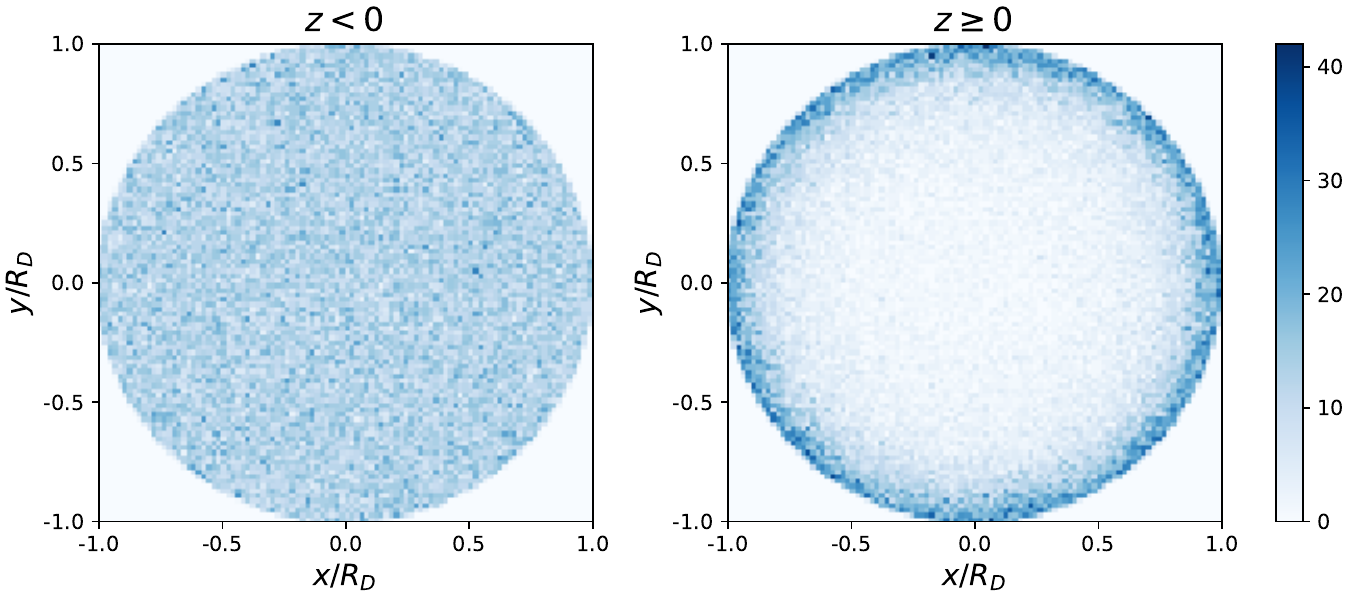}
  	\caption{``Hit  event'' distribution projected on the $x-y$ plane. We choose $m_{\chi}=50$ MeV,  $\overline{\sigma_{e}}=10^{-31}\,\mathrm{cm^{2}}$ and $v_{\chi}^{\,0}=0.02\,c$. The left panel shows the near side ($z<0$) while the right panel shows the far side ($z\geq 0$). The initial event number is $N=10^5$. Both $x$ and $y$ axes are equally divided into $100$ bins. }
  	\label{eventsxy}
\end{figure*} 
\subsection{Reconstructing the BDP flux distribution}

For each scattering, we determine the electron recoil energy $E_{R}$ and the momentum transfer $q$ using the ionization form factor. According to Eq. (\ref{DCSI}),  $q \times K(E_{R},q)$ describes the joint probability of $E_{R}$ and $q$ in an ionization process. In Fig.\,\ref{ARO}, $K(E_{R},q)$ for all elements listed in Table\,\ref{EarthModel} are demonstrated. It is worth noting that when the binding energy of an electron is much smaller than the BDP kinetic energy, $E_{R}$ and $q$ become closely correlated, and the most probable values are those satisfying $E_{R}\simeq {q^{2}}/{(2m_{\mathrm{e}}})$.
We use the generalized acceptance-rejection method \cite{arm} to acquire a $(E_{R},q)$ pair corresponding to the probability distribution given by $q \times K(E_{R},q)$.

A further dynamical constraint on the $(E_{R},q)$ pair is imposed, $E_R\leq q v_{\rm i}-q^2/(2m_\chi)$, which is equivalent to the condition of $q\in (q_{-},q_{+})$ used in Eq.\,(\ref{DCSI}). 

The magnitude of the BDP final velocity, $v_{\mathrm{f}}$, after the scattering is written as,
\begin{equation}
v_{\mathrm{f}} = \sqrt{v_{\mathrm{i}}^2-2 E_R/m_\chi} \ .
\end{equation}
The polar angle of the final velocity, $\alpha$, respect to the direction of the initial velocity can be calculated as
\begin{equation} 
v_{\mathrm{i}}^{2} + v_{\mathrm{f}}^{2} - 2v_{\mathrm{i}}v_{\mathrm{f}}\cos\alpha = \frac{q^{2}}{m_{\chi}^2}\ .
\end{equation}
To fully determine the direction of the BDP final state after the scattering, we sample the azimuthal angle $\beta$ following a uniform distribution $[0,2\pi)$.
The final state of the BDP in each iteration is thus determined, including its location $\vec{x}_{\rm f}$ and velocity $\vec{v}_{\rm f}$. These will be used as the inputs for the next iteration.

There are two conditions for the simulation to stop. First, there is a minimal recoil energy required to ionize an electron in xenon. It can be used to set a lower bound for the BDP velocity as $v_{\textrm{min}}^{\textrm{ion}} = \sqrt{2E_{R}^{\textrm{min}}/m_{\chi}}$, with $E_{R}^{\textrm{min}} \equiv 10$ eV is set in this study. Thus the threshold velocity in our simulation is chosen to be the maximum value between the DM virial velocity, i.e. $10^{-3}\,c$, and $v_{\textrm{min}}^{\textrm{ion}}$. 
 Second, the BDP may reach the surface $r = R_D$ before its velocity becomes smaller than the threshold velocity, in which case it is no longer relevant. Under both conditions, the simulation of the event will be stopped automatically.

After the simulation, we define the ``hit events'' as the ones which reach the surface  $r = R_D$. We collect the velocity and position of each event. If the BDP-electron interaction is strong, the BDP loses its energy rapidly in the Earth, and the BDP particles can only penetrate the $r = R_D$ sphere at most once, i.e. when they just enter the Earth. On the other hand, when the interaction is weak, most of the BDP particles travel across the $r = R_D$ sphere twice, this leads to a doubling of the number of events, to $2N$. 
In this subsection, we explain how to convert the distribution of ``hit events''  to the BDP velocity distribution that can be used to calculate the event rate in a detector.

In Fig.\,\ref{eventsxy},  we show an example of the ``hit  event'' distribution projected to the  $x-y$ plane in both the near side ($z < 0$) and the far side ($z \ge 0$) of the Earth. 
On the near side, the events are almost equally distributed on the $x-y$ plane, which is consistent with the initial setup in Sec. (\ref{IS}).
However on the far side, the events are more densely distributed near $\sqrt{x^2 + y^2} \simeq R_E$ where $R_E$ is the radius of the Earth. 

For the parameter space we are interested in, the transverse component of the BDP velocity is much smaller than the one along the $z-$axis after the propagation, thus the event rate in a detector can be approximately calculated using the modified BDP flux along the $z-$axis.
With a proper normalization, the ``hit  event'' rate per area on the  $r = R_D$ sphere is simply related to the modified BDP flux by a factor of $1/{\rm cos}\theta$.

\subsubsection{Reconstruction of the general velocity distribution}
In more general cases, the BDP can reach the detector from all directions. In this section, we study the conversion from the ``hit event'' rate per area to the general velocity distribution.


The number density of BDP particles with velocities within ${\rm d}^3{\vec v}_\chi$ is
\begin{equation}
n_{\chi}\ f({\vec v}_\chi)\ {\rm d}^3{\vec v}_\chi,
\end{equation}
where $n_{\chi}$ is the BDP number density and $f({\vec v}_\chi)$ is the 3-velocity distribution.
For an infinitesimal area element ${\rm d}{\vec s}$, the rate of particles hitting this area with velocities within ${\rm d}^3{\vec v}_\chi$ is 
\begin{equation}
\frac{{\rm d} h({\vec v}_\chi)}{{\rm d}^3 {\vec v}_\chi\ {\rm d}t} = n_{\chi}\  f({\vec v}_\chi)\ \left({\vec v}_\chi \cdot{\rm}{\rm d}{\vec s}\,\right).
\end{equation}
This can be used to calculate the the differential BDP event rate at a detector, 
\begin{equation}
\frac{{\rm d} R}{{\rm d} E_R} = N_d \int {\rm d}^3 {\vec v}_\chi \ n_\chi\ f({\vec v}_\chi)\ |\vec{v}_\chi|\ \frac{{\rm d}\sigma_{{\rm ion}}}{{\rm d} E_R} 
\end{equation}
where $N_d$ is the number of target atoms in the detector.

\section{Constraints on non-distorted BDP}\label{constraints}
The first results from the XENONnT experiment \cite{XENONCollaboration:2022kmb} for $E_R < 30$ keV ruled out the previous excess of the electron recoil events seen by XENON1T  \cite{XENON:2020rca}, thus providing the most stringent constraints on the BDP scenario.

The BDP model discussed in the main text contains several parameters, including the mass $m_\chi$, the incoming velocity $\vec{v}_\chi^{\,0}$, the total flux $\Phi_0$, and the cross section for scattering on free electrons $\overline{\sigma_e}$. To simplify the analysis, we consider the BDP flux without the effect of Earth's shielding. Under this assumption, the product $\overline{\sigma_e} \times \Phi_0$ is degenerate, and the incoming direction of $\vec{v}_\chi^{\,0}$ is not relevant.
The ratio of $\overline{\sigma_e}$ and $\Phi_0$ becomes significant once $\overline{\sigma_e}$ is large enough to cause shielding associated with the propagation inside the Earth, as discussed in the main text. Therefore, 
one is left with three free parameters: $\overline{\sigma_e} \times \Phi_0$, $m_\chi$ and $v_\chi^{0}$.

\begin{figure}[t!] 
    \centering
    \includegraphics[width=9cm]{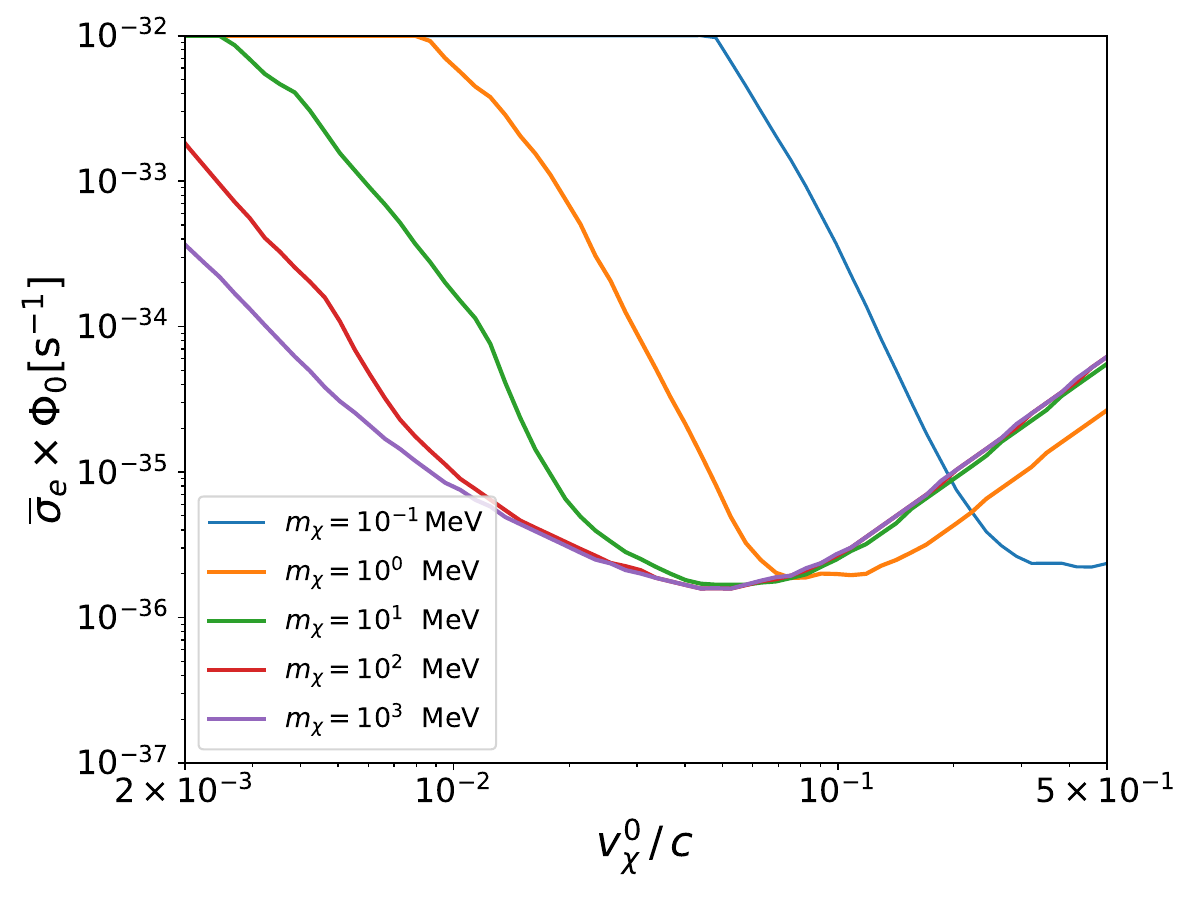}
    \caption{ The $90\%$ exclusion regions on the product $\overline{\sigma_e} \times \Phi_0$  for several values of $m_\chi$ and a range of $v_\chi^0$, plotted  using the first results of the XENONnT experiment \cite{XENONCollaboration:2022kmb}.}
    \label{fig:cnt}
\end{figure}

For each pair of $m_\chi$ and $v_\chi^{0}$, we calculate the corresponding  $-2\Delta\ln\mathcal{L}=-2\ln(\mathcal{L}_{S+B}/\mathcal{L}_{B})$ using the XENONnT data \cite{XENONCollaboration:2022kmb}, where $\mathcal{L}$ is the likelihood function, and derive the $90\%$ upper bounds on $\overline{\sigma_e} \times \Phi_0$ via requiring $-2\Delta\ln\mathcal{L}=-2\ln(\mathcal{L}_{S+B}/\mathcal{L}_{S+B}^{\rm max})=2.71$. The results are shown in Fig.\,\ref{fig:cnt}. For a BDP mass  $\sim \mathcal{O} (1)$ MeV, a large $v_\chi^0$ is required for the electron recoil energy to be above the threshold. On the other hand, with both $m_\chi$ and $v_\chi^0$ sufficiently large, the elastic scattering limit with $dR/d E_R$ scaling as $(v_\chi^{0})^2$ is recovered, as discussed in the previous appendix.
Based on the results in Fig.\,\ref{fig:cnt}, we choose our benchmark parameter in the main text to be  $\overline{\sigma_e} \times \Phi_0 = 10^{-36} {\rm s}^{-1}$.

\bibliographystyle{unsrt}

\end{document}